\documentclass[%twocolumn,
aps,nofootinbib,
showpacs,showkeys,preprint
tightenlines,preprintnumbers,] {revtex4}

\usepackage{epsf,epsfig,subfigure,graphicx,amsmath,amssymb %,mathtools
}
\usepackage{color}
\newcommand{\dis}[1]{\begin{equation}\begin{split}#1\end{split}\end{equation}}
\newcommand{\ie}{{\it i.e.~}}
\newcommand{\etal}{{\it et al.\,}}

\newcommand{\Qem}{Q_{\rm em}}

\newcommand{\tev}{\,\textrm{TeV}}
\newcommand{\gev}{\,\textrm{GeV}}

\newcommand{\Mg}{{M_{\rm GUT}}}

\newcommand{\Uanom}{U(1)$_{\rm anom}$}
\newcommand{\Ufr}{U(1)$_{\rm fr}$}

\def\sw0{{$\sin^2\theta_W^0$}}

\newcommand{\Z}{{\bf Z}}

\def\E6{{\rm E_6}}

\def\EE8{{\rm E_8\times E_8'}}

\def\flip{SU$(5)_{\rm flip}$}

\def\four{{\bf 4}}

\def\three{{\bf 3}}

\def\one{{\bf 1}}

\def\two{{\bf 2}}
\def\five{{\bf 5}}
\def\ten{{\bf 10}}
\def\tenb{{\overline{\bf 10}}}

\def\fiveb{{\overline{\bf 5}}}

\def\three{{\bf 3}}

\begin{document}

\draft

\title{\Large\bf Theory of flavors: String compactification}

\author{ Jihn E.  Kim}
\address
{Department of Physics, Kyung Hee University, 26 Gyungheedaero, Dongdaemun-Gu, Seoul 02447, Republic of Korea, and\\
Center for Axion and Precision Physics Research (Institute of Basic Science), KAIST Munji Campus, 193 Munjiro,  Daejeon 34051, Republic of Korea, and \\
  Department of Physics and Astronomy, Seoul National University, 1 Gwanakro, Gwanak-Gu, Seoul 08826, Republic of Korea 
}
 
\begin{abstract} 
We present  a calculating method for the quark and lepton mixing angles.
After a general discussion in field theoretic models,  we present a working model  from a string compactification through $\Z_{12-I}$ orbifold compactification. It is beyond presenting just three families of the standard model but is the first example from string compactification  successfully fitting to the observed data.   Assuming that all Yukawa couplings from string compactification are real,   we  also comment on a relation between the CP phases in the Jarlskog determinants obtained from the CKM and PMNS matrices. The flipped SU(5) model leads to  the doublet-triplet splitting and possible proton decay operators. It is shown that  the vacuum expectation values can be tuned such that the proton lifetime is long enough.   
 
\end{abstract}
\maketitle

%%%%%%%%%%%%%%%%%%%%%%%%%%%%%%%%%%%%%%%%%%%%%%%%%%%%  
%%%%%%%%%%%%%%%%%%%%%%%%%%%%%%%%%%%%%%%%%%%%%%%%%%%%

\section{Introduction}\label{sec:Introduction}
``How is the current allocation of flavors realized?'' is the most urgent and also interesting theoretical problem in the Standard Model (SM). Extension of the SM to grand unification (GUT) and string models \cite{Kim15Ufamily,KimKyaeNam17} continues to require to solve this flavor problem. Gauge symmetries as family groups should satisfy the anomaly freedom, which can be achieved in extended GUTs \cite{Kim80} and in models without   anomaly   \cite{King01}. Not to worry about the gauge anomalies, sometimes global symmetries are used for the family groups \cite{Reiss82,Wilczek82,Gelmini83}. It has been reviewed at several places  \cite{Fate18,Ramond85}.

In the SM, the difference of families is manifested in the Cabibbo-Kobayashi-Maskawa (CKM) matrix in the quark sector \cite{Cabibbo63,KM73} and in the Pontecorvo-Maki-Nakagawa-Sakada (PMNS) matrix in the leptonic sector\cite{PMNS1,PMNS2}.  To relate the left(L) and right(R) mixing angle parameters, the flavor group $G_f$ has been introduced  to obtain more relations between flavor parameters \cite{Weinberg77, Fritzsch78, WilZee77,Nielsen79}. In most cases, a factor flavor group $G_f$ is introduced in addition to the SM or GUT. On the other hand, an attractive mechanism is to unify all the fermion representations in an irreducible set of SU($N$) representations of an extended GUT \cite{Georgi79,Frampton79,FramptonPRL79, Kim80}. The $\EE8$ gauge group can be considered to belong to this class but in ten dimensions. So, compactification of six extra dimensions may be the key to the unification of families in ten dimensional superstring models.

A notable difference between the CKM and the  PMNS matrices lies in the fact that in the CKM matrix the large elements are located in the diagonal entries while it is not so in the PMNS matrix. So, for the CKM parameters the quark mass ratios were used before \cite{Weinberg77, Fritzsch78}. On the other hand, for the PMNS parameters non-Abelian discrete groups   are used \cite{Harrison95,Ma01}. One may say that there is one similarity in the CP phases of the CKM and PMNS matrices. The CKM phase is close to $90$ degrees in the Kim-Seo (KS) parametrization \cite{KimSeo11} and the PMNS phase is $-90$ degrees (but with a large error bars) \cite{T2K17}. Even, there exists an attempt to unify these CP phases
\cite{NamCKMPMNS}. 

To reduce the number of parameters in the flavor sector,  family symmetries can be used. Simple ones are U(1) groups. But, to introduce a hierarchy, vacuum expectation values (VEVs) of the SM singlets are suggested, which is known as the Froggatt-Nielsen (FN) mechanism \cite{FN79}.
   
In this paper, we study singlet representations beyond the SM based on   family symmetry groups.  For various reasons in field theoretic models,  we consider U(1)$^2$ among which one is anomalous and the other is anomaly free.  We attempt to obtain singlets  from the orbifold compactification of the $\EE8$ heterotic string \cite{Gross85} based on the simplest $\Z_{12-I}$ lattice  \cite{LNP696,KimKyae07}.  Fixed points of 13 prime orbifolds listed in \cite{Kaplunovsky90} shows that the $\Z_{12-I}$ lattice can be considered to be the simplest because there are only three fixed points.\footnote{The $\Z_3$ orbifold, seemingly looking very simple, has 27 fixed points. Being simple, $\Z_{12-I}$ may not be general enough, but can present the basic working principles in terms of small number of fields.}

In Sec. \ref{sec:Fmasses}, we briefly recapitulate the fermion mass structure: Dirac fermions of charged leptons and quarks, and Majorana fermions for the SM neutrinos. We set up the scheme to use Weyl fermions to express both the Dirac and Majorana masses pressented in subsequent sections. Those who are familiar to these can skip this section.
In Sec. \ref{sec:twoU1s},    we present a beyond-SM  with two U(1)'s  toward useful fermion mass textures, where one is \Uanom~global symmetry for the ``invisible'' axion and the other is anomaly free gauged U(1).
In Sec. \ref{sec:string},  we obtain a successful flavor structure from $\Z_{12-I}$ orbifold compactification. The 3rd family  is assumed to be the one from the untwisted sector $U$.
Sec. \ref{sec:Conclusion} is a conclusion.  
  
 %%%%%
\section{Fermion masses}\label{sec:Fmasses}
 
For continuous parameters of transformation, let us begin with the axial-vector currents of fermions
\dis{
 J^\mu_{\Gamma}= \bar{\Psi}\gamma^\mu \gamma_5 \Gamma \Psi
}
where $\Gamma$ is a charge operator and $\Psi$ is a column vector of three fermions (families).
The divergence of the current is
\dis{
\partial_\mu J^\mu_{\Gamma}=\frac{A_I}{32\pi^2}G^I\tilde{G}^I
+2\mu J^5_{\Gamma}
}
where $A_I$ is the anomaly coefficient of the gauge fields $A_\mu^I$ for the gauge group SU($N$)$_I$, and $J^5_{\Gamma}$  depends on the masses of fermions,
\dis{
 J^5_{\Gamma}=\frac{1}{\mu}\bar{\Psi}i\gamma_5 M_\Gamma \Psi ,\label{eq:MassM}
}
where $\mu$ is a mass scale and $M$ is the mass matrix in the flavor basis. The anomaly term is a flavor singlet which can be written in terms of a flavor singlet quark fields in the $\theta$-vacuum, e.g. for two flavors in SU(3)$_c$
 \cite{KimPRP87},
\dis{
G^I\tilde{G}^I\propto  \sqrt{\frac{m_um_d}{2\cos\theta+(1+Z^2)/Z}}~(\bar{u}i\gamma_5 u+\bar{d}i\gamma_5 d)\sin\theta
} 
where $Z=m_u/m_d\simeq\frac12$. Obviously, the anomalous and anomaly free terms give nonzero trace for the fermion mass matrix. In Ref. \cite{Ramond85}, two U(1) symmetries were considered, one anomalous and the other anomaly-free. The anomalous global symmetry is to introduce the so-called ``invisible'' axion. Since the sum of quark masses is nonzero and large O($m_t$), we also attempt to have the anomalous U(1). The anomalous U(1) must be a global symmetry and the anomaly free part can be a gauge symmetry.  Let us start using two component fermions to write down mass terms.

%%%%%%%
\subsection{Weyl fermions}
A four component Dirac spinor, e.g. for the electron field, can be split into two Weyl spinors $\xi$ and $\eta$,  
\dis{
\psi_e=\begin{pmatrix} e_L\\ e_R
\end{pmatrix}=\begin{pmatrix} e_{L1}\\  e_{L2}\\ e_{R1}\\  e_{R2}
\end{pmatrix}\to  \begin{pmatrix} \xi_L\\ \eta_R
\end{pmatrix}.
}
Gauge interactions do not change the chirality. Quantum fild $e_L$  destroys a L-handed electron and  creates a R-handed positron. But,  $e_L$  has nothing to do with destroying  a R-handed electron $e_R$  and creating a L-handed positron. On the other hand, the anti-particle of the L-handed electron $e_L=(e_{L1},e_{L2},0,0)^T$ is
\dis{
(e_L)^c= \begin{pmatrix}e_{L1}\\  e_{L2}\\ 0\\ 0 \end{pmatrix}^c=\left(\frac{1+\gamma_5}{2}e \right)^c=i\sigma_2 \left(\frac{1+\gamma_5^*}{2}e^*\right) 
=\frac{1-\gamma_5}{2}  \begin{pmatrix}0\\ 0\\ e_{L2}^* \\ -e_{L1}^* \end{pmatrix}
}
which is a R-handed field. This R-handed field destroys R-handed positron and creates L-handed electron. With these two Weyl fields, we can destroy L- and R-electrons and   create  L- and   R-positrons, which is done by a four-component Direc electron. Thus, two Weyl fields are enough at this stage.   With the Weyl field $\xi$, let us construct a Lorentz invariant $\epsilon_{ij}e_{Li}e_{Lj}$. It is the mass term but the electron number is broken by this term. So, for charged particles, one Weyl field cannot be massive. For neutrinos, one Weyl field can give a mass term which is known to be Majorana mass. In this paper, we will use Weyl fields even for expressing Majorana masses,
\dis{
\epsilon_{ij}(\xi^T)^i\xi^j=\xi^T\begin{pmatrix}0&1\\  -1&0
\end{pmatrix}\xi=\xi^T i\sigma_2\xi=\xi^T \gamma_1\gamma_3\xi.\label{eq:MajMass}
}
For a Dirac mass we use the opposite chirality, \ie
\dis{
\xi^{Ti}_L=\epsilon^{ik}\overline{\eta}_{k R}\to \epsilon_{ij} \epsilon^{jk}
\overline{\eta}_{k R}=\overline{\eta}_{i R}
}
so that (\ref{eq:MajMass}) becomes
\dis{
\overline{\eta}_{R}\xi _L.\label{eq:DirMass}
}
Assigning the same charge conjugation for $\xi$ fields in
(\ref{eq:MajMass}), the Majorana mass term breaks $C$, but the Dirac mass term (\ref{eq:DirMass}) can preserve $C$ by assigning the same $C$'s for $\xi$ and $\eta$.  Discussing both Majorana and Dirac masses,  using the Weyl fermion is therefore simple enough.

%%%%%%%
\subsection{$m_b\simeq m_\tau$ and Georgi-Jarlskog relation}\label{subsec:GJrel}

The observed ratio of the third family masses $m_b/m_\tau\simeq 4.5/1.5\approx 3$ hints that $m_b\simeq m_\tau$ at the unification scale. The factor 3 arises by renormalization group evolution \cite{GQW74}. In the Georgi-Glashow (GG) SU(5) \cite{GG74}, $\tenb_f\,\five_f\,\five_H$ gives the same mass to $b$ and $\tau$ by $\langle\five_H\rangle$ and it is considered to be a success of the GUT \cite{Buras78}. For the muon and strange quark, however, there is a  big problem in the GG SU(5) model. The   low energy mass ratio at 100 MeV is $m_s/m_\mu\approx 1$ while the renormalization group evolution expects it to be 3 if $m_s\simeq m_\mu$ at the unification scale. If $m_s/m_\mu\simeq\frac13$ at the unification scale, then the low energy mass ratio is understandable. But, this is a big problem  with Higgs quintets only. One way out is the Georgi-Jarlskog relation introducing a big Higgs representation  ${\bf 45}_H$   \cite{GJ79}.  If $\langle {\bf 45}_H\rangle$ is the leading contribution to the second family fermions in the GG model, then   $m_s/m_\mu\simeq\frac13$ is obtained at the unification scale. To present a rationale for   ${\bf 45}_H$ for the needed mass matrix texture, two U(1)'s were suggested long time ago \cite{Ramond85}.

%%%%%%%
\subsection{Flipped SU(5)}\label{subsec:flip}

Our terminology of flipped SU(5) is a rank 5 gauge group \flip=SU(5)$\times$U(1)$_X$. Representations will be denoted as SU(5)$_{\rm U(1)}$. In the flipped SU(5) \cite{Barr82,DKN84}, masses of charged leptons and $d$-type quarks are not related, which is considered to be a merit  in relating masses.
In string compactification, reasonable supersymmetric SM's are obtained from compactification of heterotic string. The reason is the following. For $N=1$ supersymmetric (SUSY) massless fields, only the completely antisymmetric representations are allowed with one compactification scale from heterotic string \cite{LNP696}.  If the Higgs fields breaking a GUT group appear as massless spectra, then there is no adjoint representation at the GUT scale which is needed for breaking the GG SU(5) or SO(10) GUT \cite{SO10GUT1,SO10GUT2} or some Pati-Salam (PS) \cite{PS73} gauge groups.\footnote{The electroweak PS gauge group SU(2)$_L\times$SU(2)$_R\times$U(1)$_{B-L}$ is broken by a GUT scale VEV to  SM$\times$U(1)$_{B-L}$, needing an adjoint representation not to reduce the rank. Usually, it is denoted as $\Delta=(\one,\three,0)$. A VEV of an adjoint representation does not reduce the rank of the gauge group. But, note that an adjoint representation is possible in some scenarios in $\Z_{6-II}$ by introducing two compactification scales for  $N=2$ SUSY in an interim effective 5 dimensions \cite{Raby05}.} 
In \flip, the representation $\ten_{+1}\oplus \tenb_{-1}$ can break the rank-5 \flip~down to the rank-4 SM gauge group.
At the GUT level, therefore only the flipped SU(5) is actually realized in several string compactifications \cite{Ellis89,KimKyae07,Huh09,Kim15Ufamily}.

 %%%%%
\section{\Uanom$\times$\Ufr~ family symmetry}\label{sec:twoU1s}

We introduce supersymmetry and two U(1) gauge symmetries, \Uanom$\times$\Ufr, where \Uanom~is anomalous and \Ufr~is free of gauge anomalies.   Dangerous dimension-4 superpotential of the 1st family members triggering proton decay is  
\dis{
q_1q_1q_1l_1
}
where the subscrit 1 denotes the first family. U(1)$_{B-L}$ allows the above superpotential but \Uanom~or \Ufr~may not allow it. Thus, the extra U(1)'s may be useful forbidding some unwanted proton decay operators. In string compactification, one has to check the \Uanom$\times$\Ufr~quantum numbers of the first family members to see if the unwanted proton decay operators are forbidden. If the proton decay problem is safe, one can consider the superpotentials generating fermion masses.

The mass eigenstates of quarks, $q^m$, are related to the weak eigenstates by L- and R-unitary matrices, $U$ and $V$,
\dis{
&q_{d,u\,L}^m=U_{d,u}q_{d,u\,L}^w\\
&q_{d,u\,R}^m=V_{d,u}q_{d,u\,R}^w,
}
and the charged $W^+_\mu$ coupling for the L-handed quark doublets is
\dis{
W=U^\dagger_u U_d,\label{eq:CKMdef}
}
which is the CKM matrix. 

%%%%%%%
\subsection{Effects of   \Uanom~ on the texture of mass matrix}
To see the essence, let us consider two families of quarks. Let us choose the basis where $\Qem=+\frac23$ quarks are already mass eigenstates. Then, the mass matrices of  weak and mass eigenstates of  $\Qem=-\frac13$ quarks are related by $M^w_d=V_d^\dagger M_d^mU_d$. Parametrizing the unitary matrices as
\dis{
U_d=\begin{pmatrix}c_1& s_1\\ -s_1&c_1 \end{pmatrix},~~V_d=\begin{pmatrix}c_2& s_2\\ -s_2&c_2 \end{pmatrix}
}
where $c_i=\cos\theta_i$ and  $s_i=\sin\theta_i$ for $i=1,2$. Thus, $M^w_d$ is given by
\dis{
M^w_d=\begin{pmatrix} c_1c_2m_d +s_1s_2m_s,& s_1c_2m_d- c_1s_2 m_s\\[0.3em] c_1s_2m_d-s_1c_2m_s,&s_1s_2m_d+c_1c_2m_s \end{pmatrix},\label{eq:Mass2w}
} 
where $m_d$ and $m_s$ are eigenvalues of the mass matrix.
If any one element of Eq. (\ref{eq:Mass2w}) is zero, then $\theta_1$ and $\theta_2$ are related. Weinberg's choice \cite{Weinberg77} is $m_d\to -m_d$ and $(M^w_d)_{11}=0$, leading to $s_1s_2/c_1c_2=m_d/m_s$. Since $\sin\theta_C\simeq \sqrt{m_d/m_s}$ numerically, we use the freedom in $V$ and choose $s_2/c_2=s_1/c_1$, which means that the R-handed fields transform in the same way as the L-handed fields. This implies that under any extra U(1) gauge group the gauge transformations of the L- and R-handed fields are identical. Thus, extra U(1)'s should be free of gauge anomalies. Therefore, if we do not consider extra quarks beyond the SM the Fritzsch texture \cite{Fritzsch78}, following Ref. \cite{Weinberg77}, is not valid with the \Uanom~gauge group. So, it is appropriate to introduce heavy quarks  to have \Uanom~together with the  Fritzsch texture.

Presence of \Uanom~gauge group requires a difference between $U_d$ and $V_d$. To reduce one more parameter, Wilczek and Zee \cite{WilZee77} choose $(M^w_d)_{12}=0$ with $s_2/c_2\simeq (s_1/c_1)^3$, which is consistent with the presence of \Uanom~gauge group.
Namely, in the presence of \Uanom~gauge group, we must choose $V$ differently from $U$ even for three families.

Similarly, let us consider two families of leptons where charged lepton mass matrix is already diagonalized.  Then, the mass matrices of  weak and mass eigenstates of Majorana neutrinos are related by $M^w_\nu=U_\nu^T  M_\nu^mU_\nu$.  Thus,
$M^w_\nu$ is given by
\dis{
M^w_\nu=\begin{pmatrix} C_1^2m_{\nu_e}+S_1^2m_{\nu_\mu},& -C_1S_1(m_{\nu_\mu}-m_{\nu_e})\\[0.5em]  -C_1S_1(m_{\nu_\mu}-m_{\nu_e}),&S_1^2m_{\nu_e}+C_1^2m_{\nu_\mu} \end{pmatrix},\label{eq:Mass2wNu}
} 
where $C_1=\cos\Theta_1$ and $S_1=\sin\Theta_1$. Since the mixing angle of the second and the third neutrinos is large, we can approximate $C_1\simeq S_1=1/\sqrt{2}$. In this case, the mass matrix is of the form\footnote{This case with two parameters is including the possibility of family indices carried by Higgs fields. If family indices  of Higgs fields are independent from the family index of quark and leptons, then there must be one parameter.}
\dis{
M^w_\nu=\begin{pmatrix} A,& -B\\[0.3em]  -B,&A \end{pmatrix}
}
where $A=(m_{\nu_\mu}+m_{\nu_e})/2$ and  $B=(m_{\nu_\mu}-m_{\nu_e})/2$, which has the permutation symmetry $S_2$ between the second and the third family indices. The useful discrete symmetries of \cite{Harrison95,Ma01} contain this $S_2$ as a subgroup.  In this case of introducing \Uanom, where we introduced only L-handed neutrinos, the anomaly freedom must be satisfied by the quantum numbers of the first family leptons or by heavy leptons.

%%%%%%%
\subsection{Quark mass matrices}\label{subsec:quarks}
%%%%%%%%%
\begin{table}[h!]
\begin{center}
\begin{tabular}{@{}cccccccc@{}} \toprule
  &  ~$\overline{u}_{1R}$~ &  ~$\overline{u}_{2R}$~ &
~ $\overline{u}_{3R}$~ &  ~$u_{1L}$~&  ~$u_{2L}$~& ~ $u_{3L}$~& ~ $\sigma$~ \\ \colrule
$Q_{\rm anom}$  &  $+2$ &  $+1$ & $0$ & $+4$ & $+2$ & $0$ & $-3$     \\ \botrule
\end{tabular} \label{tab:Qupquarks} 
\end{center}
\caption{Charges of up type quarks.}
\end{table}
%%%%%%%%%%%%%%

Let us begin with the diagonalized Dirac masses of the form (\ref{eq:DirMass}) for $\Qem=+\frac23$ quarks,  
\dis{
&\qquad\quad~\xi_{1L}\,~\xi_{2L}\,~\xi_{3L}\\
M^{\rm diag}_u=&\begin{array}{c}
\overline{\eta}_{1R}\\[0.5em]
\overline{\eta}_{2R}\\[0.5em]
\overline{\eta}_{3R}
\end{array} 
\begin{pmatrix}
m_u & 0& 0\\[0.5em]
0& m_c & 0\\[0.5em]
0 & 0 & m_t
\end{pmatrix},\label{eq:MuDiag}
}
where $\overline{\eta}_{R}=\overline{q}_{uR}^m$ and $\xi_L=q_{uL}^m$.
The diagonal form (\ref{eq:MuDiag}) with the needed hierarchy can be obtained by the U(1) charges of  Table I,
\dis{
 M_u\propto
\begin{pmatrix}
\quad \sigma^2\quad   &   0\quad &\quad~~ 0\quad~~ \\[0.5em]
0&   \sigma  \quad & 0\\[0.5em]
0 & 0\quad & 1
\end{pmatrix},
}
where $\propto\sigma $ is a SM singlet field  carrying $Q=-3$. The mass term for up-type quarks  is
\dis{
\overline{q}_{uR}^mM^{\rm diag}_uq_{uL}^m =\overline{q}_{uR}^w V_u^\dagger M^{\rm diag}_u U_uq_{uL}^w .
}
Of course, $V_u=U_u=\one$.

The mass matrix for $\Qem=-\frac13$ quarks is
\dis{
\overline{q}_{dR}^mM^{\rm diag}_dq_{dL}^m =\overline{q}_{dR}^w V_d^\dagger M^{\rm diag}_d U_dq_{dL}^w,
}
with
\dis{
M_d= V_d^\dagger \begin{pmatrix}
\quad m_d\quad   &   0  &\quad~~ 0\quad~~ \\[0.5em]
0&   m_s   & 0\\[0.5em]
0 & 0 & m_b
\end{pmatrix} 
 U_d.\label{eq:MdDiag}
}
In this paper, we use the KS parametrization \cite{KimSeo11} of the CKM matrix $W_{KS}$ where $\delta_{\rm CKM}=\frac{\pi}{2}$ is simple,\footnote{As stressed in  \cite{KimSeo11}, the CP phase in the CKM matrix is close to 90 degrees if we parametrize it by  $W_{KS}$. The phase $\delta$ is the phase in the  Jarlskog determinant \cite{Jarlskog85}.}
 \dis{
W_{KS}=U_d=
\begin{pmatrix}
\quad c_1,\quad   &   +s_1c_3,\quad &\quad~~+s_1s_3\quad~~ \\[0.5em]
-c_2s_1,& +c_1c_2c_3+s_2s_3e^{-i\delta}, \quad &+c_1c_2s_3  -s_2c_3e^{-i\delta}\\[0.5em]
 -s_1s_2e^{i\delta}, &-c_2s_3 +c_1s_2c_3 e^{i\delta}, \quad & +c_2c_3+c_1s_2s_3e^{i\delta}
\end{pmatrix},\label{eq:KSform}
}
with the unitary matrix for R-handed fields in the diagonalization process parametrized by another 4 parameters  
\dis{
V_d^\dagger=
\begin{pmatrix}
\quad c_4,\quad   &  -c_5s_4, \quad &\quad~~\quad-s_4s_5e^{-i\Delta}~~ \\[0.5em]
+s_4c_6,& +c_4c_5c_6+s_5s_6e^{i\Delta}, \quad &-c_5s_6 +c_4s_5c_6 e^{-i\Delta}\\[0.5em]
 +s_4s_6, &+c_4c_5s_6  -s_5c_6e^{i\Delta}, \quad & +c_5c_6+c_4s_5s_6e^{-i\Delta}
\end{pmatrix}.
}
Change the sign $m_d\to -m_d$, and to reduce the number of parameters let us  choose parameters of R-fields as

Then, we obtain $V_d^\dagger M^{\rm diag}_d U_d$ as 
\dis{
\begin{pmatrix}
\begin{array}{l} c_4c_1 m_d+c_5s_4c_2s_1m_s\\
+s_4s_5s_1s_2e^{-i\Delta+i\delta} m_b\\ ,\end{array}   &\begin{array}{l}  c_4s_1c_3m_d -c_5s_4c_1c_2c_3m_s\\
-c_5s_4s_2s_3e^{-i\delta}m_s+s_4s_5c_2s_3e^{-i\Delta} m_b\\
-s_4s_5c_1s_2c_3e^{-i\Delta+i\delta} m_b,
\end{array} &\begin{array}{l}c_4s_1s_3 m_d   -c_5s_4c_1c_2s_3m_s\\ +c_5s_4s_2c_3e^{-i\delta}m_s  
-s_4s_5c_2c_3e^{-i\Delta} m_b\\
-s_4s_5c_1s_2s_3e^{-i\Delta+i\delta} m_b,\end{array}  \\[1em]
&&\\
\begin{array}{l} s_4c_6c_1 m_d-c_4c_5c_6 c_2s_1m_s\\
-s_5s_6c_2s_1 e^{i\Delta}m_s\\ 
+c_5s_6s_1s_2 e^{i\delta}m_b \\
-c_4s_5c_6 s_1s_2e^{-i\Delta+i\delta}m_b\\ \\ ,\end{array}  & 
\begin{array}{l} s_4c_6s_1c_3 m_d+c_4c_5c_6 c_1c_2c_3m_s\\
+c_4c_5c_6 s_2s_3e^{-i\delta}m_s\\
+s_5s_6 c_1c_2c_3e^{i\Delta}m_s+s_5s_6 s_2s_3e^{i\Delta-i\delta}m_s \\ 
+c_5s_6c_2s_3 m_b - c_5s_6c_1s_2c_3e^{i\delta}m_b\\
-c_4s_5c_6 c_2s_3e^{-i\Delta}m_b\\
+c_4s_5c_6 c_1s_2c_3e^{-i\Delta+i\delta}m_b,\end{array}  
& 
\begin{array}{l} s_4c_6s_1s_3 m_d+c_4c_5c_6 c_1c_2s_3m_s\\
-c_4c_5c_6 s_2c_3e^{-i\delta}m_s\\
+s_5s_6 c_1c_2s_3e^{i\Delta}m_s-s_5s_6 s_2c_3e^{i\Delta-i\delta}m_s \\ 
-c_5s_6c_2c_3 m_b - c_5s_6c_1s_2s_3e^{i\delta}m_b\\
+c_4s_5c_6 c_2c_3e^{-i\Delta}m_b\\
+c_4s_5c_6 c_1s_2s_3e^{-i\Delta+i\delta}m_b,\end{array}  \\[1em]
&&\\
 \begin{array}{l} s_4s_6c_1 m_d-c_4c_5s_6 c_2s_1m_s\\
+s_5c_6c_2s_1 e^{i\Delta}m_s\\ 
-c_5c_6s_1s_2 e^{i\delta}m_b \\
-c_4s_5s_6 s_1s_2e^{-i\Delta+i\delta}m_b\\ \\ ,\end{array}   
&\begin{array}{l} s_4s_6s_1c_3 m_d+c_4c_5s_6 c_1c_2c_3m_s\\
+c_4c_5s_6 s_2s_3e^{-i\delta}m_s\\
-s_5c_6 c_1c_2c_3e^{i\Delta}m_s-s_5c_6 s_2s_3e^{i\Delta-i\delta}m_s \\ 
-c_5c_6c_2s_3 m_b + c_5c_6c_1s_2c_3e^{i\delta}m_b\\
-c_4s_5s_6 c_2s_3e^{-i\Delta}m_b\\
+c_4s_5s_6 c_1s_2c_3e^{-i\Delta+i\delta}m_b,\end{array}  
& \begin{array}{l} s_4s_6s_1s_3 m_d+c_4c_5s_6 c_1c_2s_3m_s\\
-c_4c_5s_6 s_2c_3e^{-i\delta}m_s\\
-s_5c_6 c_1c_2s_3e^{i\Delta}m_s+s_5c_6 s_2c_3e^{i\Delta-i\delta}m_s \\ 
+c_5c_6c_2c_3 m_b+ c_5c_6c_1s_2s_3e^{i\delta}m_b\\
+c_4s_5s_6 c_2c_3e^{-i\Delta}m_b\\
+c_4s_5s_6 c_1s_2s_3e^{-i\Delta+i\delta}m_b,\end{array} 
\end{pmatrix}\label{eq:Wplus}
}
Change the sign $m_d\to -m_d$, and to reduce the number of parameters let us  choose parameters of R-fields as
\dis{
\frac{s_4}{c_4}=\frac{s_1}{c_1} , ~\frac{s_5}{c_5}=\frac{m_s}{m_b}\frac{s_2}{c_2},~ s_6=0,~ \Delta=\delta.  
}
Note that $s_4\gg s_1s_2^2$. Then,  keeping the largest terms in the weak basis mass matrix, 
\dis{
M_d^w &\simeq \begin{pmatrix}
0  &\begin{array}{l}   -s_1c_5c_1c_2^{-1}c_3m_s ,
\end{array} &\begin{array}{l}   -c_5c_1s_1c_2^{-1}s_3m_s ,\end{array}  
\\
\begin{array}{l}   
-c_1c_5  s_1 c_2^{-1} m_s ,\end{array}  & 
\begin{array}{l}  +c_1c_5  c_1c_3 c_2^{-1} m_s,\end{array}  
& 
\begin{array}{l}  
+c_4c_5  c_1 s_3 c_2^{-1} m_s,\end{array}  
&&\\
 \begin{array}{l}   
-c_5 s_1s_2 e^{i\delta}m_b  ,\end{array}   
&\begin{array}{l}    
-c_5 c_2s_3 m_b + c_5 c_1s_2c_3e^{i\delta}m_b ,\end{array}  
& \begin{array}{l}   
+c_5 c_2c_3 m_b 
 ,\end{array} 
\end{pmatrix}\\[0.5em]
&\simeq  \frac{1}{c_5 c_2c_3m_b}\begin{pmatrix}
0, &-4.43\times 10^{-3}, & -0.690\times 10^{-4}\\
-4.43 \times 10^{-3}, &  1.918\times 10^{-2},&2.99\times 10^{-4} \\
-0.9008\times 10^{-2}e^{i\delta}, &-1.557\times 10^{-2}+3.90\times 10^{-2}e^{i\delta},& 1
\end{pmatrix}\label{eq:dMass}
}
where we used
\dis{
&m_s=93.8\, {\rm MeV} ,~ m_b= 4.65\, {\rm GeV},~ \frac{m_s}{m_b}=0.0202  ,\\
&s_1=0.2252, c_1=0.9743, s_2=0.0400, c_2=0.9992, s_3=0.01557, c_3=0.9999,\\
&\frac{s_5}{c_5}= 0.809\times 10^{-5},~\frac{s_4}{c_4}=\frac{s_1}{c_1}.
\label{eq:input}
}
In Eq. (\ref{eq:dMass}), the (32) element can be $4.2\times 10^{-2}e^{i\delta'}$ where $\tan(\pi-\delta')=-0.9286\sin\delta$. The Yukawa couplings run from the compactification  scale down to the electroweak scale in which case the  dimensionless Yukawa couplings  cannot be used directly for assigning the input mass parameters. But all Yukawa couplings leading to parameters in Eq. (\ref{eq:dMass}) are arising from the VEVs of FN singlet fields and we may use those given in Eq. (\ref{eq:dMass}) as the input parameters determining the CKM matrix.
 
%%%%%%%%
\subsubsection{Mass matrix in field theory}

Let us present a possibility of obtaining a mass matrix similar to Eq. (\ref{eq:dMass}) in field theory. Let the \Uanom$\times$\Ufr~quantum numbers are shown as $(Q_{\rm anom},Q_{\rm fr})$. After diagonalizing the $\Qem=\frac23$ quark masses, the L- and R-fields of  $\Qem=-\frac13$ quarks, $\xi$ and $\overline{\eta}$, 
 quantum numbers of $\overline{\eta}\xi$ are
\dis{
&\qquad \qquad\quad\quad~\xi_{1L}(2,-1)\quad~\xi_{2L}(0,1)\quad~ \xi_{3L}(1,0)\\
&\begin{array}{c}
\overline{\eta}_{1R}(1,-1)\\[0.5em]
\overline{\eta}_{2R}(2,1)\\[0.5em]
\overline{\eta}_{3R}(-1,0)
\end{array} 
\begin{pmatrix}
\quad (3,-2)\quad  &\quad (1,0)\quad &\quad (2,-1) \quad\\[0.5em]
(4,0)& (2,2) & \quad(3,1) \quad\\[0.5em]
\quad(1,-1) \quad&(-1,1) &\quad(0,0) \quad 
\end{pmatrix},\label{eq:quanQLQR}
}
Thus, the quantum numbers of Higgs fields appearing in the mass matrix are 
\dis{
Q(M^d)=
\begin{pmatrix}
\quad (-3,+2)\quad  &\quad (-1,0)\quad &\quad (-2,+1) \quad\\[0.5em]
(-4,0)& (-2,-2) & \quad(-3,-1) \quad\\[0.5em]
\quad(-1,+1) \quad&(+1,-1) &\quad(0,0) \quad 
\end{pmatrix}.\label{eq:Quansigma}
}
To mimick the order appearing in Eq. (\ref{eq:dMass}), let us introduce small parameters via the FN SM singlet fields, $\delta_1,\delta_2,\delta_3,\Delta_1,\Delta_2, \epsilon_1$, and $ \epsilon_2$ whose quantum numbers are shown in Table II. 
%%%%%%%%%
\begin{table}[h!]
\begin{center}
\begin{tabular}{@{}cccccccc@{}} \toprule
FN fields &  ~$\delta_1$~ &  ~$\delta_2$~ &
~ $\delta_3$~ &  ~$\Delta_1$~&  ~$\Delta_2$~& ~ $\epsilon_1$~& ~ $\epsilon_2$~  \\ \colrule
$Q_{\rm anom}$  &  $-1$ &  $-2$ & $+1$ & $-4$ & $-2$ & $0$ &  $0$ \\
$Q_{\rm fr}$  & $0$ &  $0$ & $0$ & $0$    & $-2$  & $+1$    & $-1$     \\ \botrule
\end{tabular} \label{tab:FN1} 
\end{center}
\caption{FN singlet fields}
\end{table}
%%%%%%%%%%%%%%

Let us assume that only $\delta_1$ and $\Delta_1$  have  complex VEV's, $\delta_1e^{i\delta}$ and  $\Delta_1e^{i\Delta}$ while all the other FN fields have real VEV's. Thus, $M^d$ can be written as 
\dis{
M^d=m_b\begin{pmatrix}
\quad  |\delta_1|^3e^{3i\delta}\epsilon_1^{2} \quad
  &~-|\delta_1|e^{i\delta}\quad &~  \delta_2\epsilon_1,\Delta_2\epsilon_1^3  \quad\\[0.5em]
-|\Delta_1|e^{i\Delta}& -|\Delta_2|   & ~|\Delta_1|e^{i\Delta}\delta_3\epsilon_2,\Delta_2|\delta_1|e^{i\delta}\epsilon_1 \quad
\\[0.5em]
-|\delta_1| e^{i\delta}\epsilon_1&(|\delta_1|e^{-i\delta} +\delta_3  )\epsilon_2  &~1 \quad
\end{pmatrix},\label{eq:TreeM0}
} 
where the overal constant is $m_b$ and for simplicity we do not write group theoretic numbers of  O(1). The element $M^d_{23}$ can have  $\delta_1^3e^{3i\delta}\epsilon_2  $ which we neglected because it is much smaller than the other terms. A negative signed phase in $M^d_{32}$ of Eq. (\ref{eq:TreeM0}) may need a complex conjugated field, but we do not introduce complex conjugated fields in the mass matrix  for  a SUSY extension.

%%%%%%%
\subsection{With SUSY}\label{subsec:SUSY}

Not to introduce complex conjugated fields in the mass matrix, let us consider the fields presented in Table III,
%%%%%%%%%
\begin{table}[t!]
\begin{center}
\begin{tabular}{@{}cccccccc@{}} \toprule
FN fields &  ~$\delta_1$~  &
~ $\delta_3$~ &  ~$\Delta_1$~&  ~$\Delta_2$&  ~$\Delta_3$~& ~ $\epsilon_1$~& ~ $\epsilon_2$~  \\ \colrule
$Q_{\rm anom}$  &  $-1$   & $+1$ & $-4$ & $-2$& $-1$ & $0$ &  $0$ \\
$Q_{\rm fr}$  & $0$   & $0$ & $0$    & $-2$   & $+1$  & $+1$    & $-1$     \\ \botrule
\end{tabular} \label{FNsusy} 
\end{center}
\caption{L-handed chiral fields for SUSY extension.}
\end{table}
%%%%%%%%%%%%%%

\dis{
M^d=m_b\begin{pmatrix}
\quad  |\delta_1|^3e^{3i\delta}\epsilon_1^{2},\Delta_2 |\Delta_3|e^{i\Delta_{\rm ph3}}  \epsilon_1^3\quad
  &~-|\delta_1|e^{i\delta}, |\Delta_3|e^{i\Delta_{\rm ph3}} \epsilon_2\quad &~  \delta_1\Delta_3,\Delta_2\epsilon_1^3  \quad\\[0.5em]
-|\Delta_1|e^{i\Delta_{\rm ph1}} & -|\Delta_2|  & ~|\Delta_1|e^{i\Delta_{\rm ph1}}\delta_3\epsilon_2,\Delta_2|\delta_1|e^{i\delta}\epsilon_1 ,\Delta_2 |\Delta_3|e^{i\Delta_{\rm ph3}} \quad
\\[0.5em]
-|\delta_1| e^{i\delta}\epsilon_1 & -|\Delta_3|e^{i\Delta_3}  ,-\delta_3  \epsilon_2  &~1 \quad
\end{pmatrix},\label{eq:TreeM} 
} 
For  $\delta_3=O(1)$ and  small $\Delta_3$ and $\epsilon_2$, and redefine $\xi_2\to \xi_2 e^{-i\delta},\overline{\eta}_2\to \overline{\eta}_2 e^{-i\Delta_1}$. 
By choosing $\Delta_{\rm ph1}=\Delta_{\rm ph3}= \delta$, and $\delta=\frac{\pi}{2}$, we obtain
\dis{
M^d\simeq m_b\begin{pmatrix}
\quad  \mp|\delta_1|^3 \epsilon_1^{2}\, i \quad
  &~-|\delta_1|\quad &~  -a_1|\Delta_2|\epsilon_1^3  \quad\\[0.5em]
-|\Delta_1|&|\Delta_2 |  & ~a_2|\Delta_1| \delta_3\epsilon_2+a_3\Delta_2|\delta_1| \epsilon_1+a_4\Delta_2 |\Delta_3|   \quad
\\[0.5em]
-|\delta_1|  \epsilon_1 \, i& -|\Delta_3| +\delta_3  \epsilon_2\, i  &~1 \quad
\end{pmatrix},\label{eq:TreeM}
}  
where we introduced O(1) numbers $a_{1,2,3,4}$.
 Firstly, $|\Delta_1|=|\delta_1|=4.432\times 10^{-3}$ and require $a_1 \Delta_2\epsilon_1^3=0.690\times 10^{-4}$ (with $a_1\simeq 1$). Let $|\Delta_2|=0.01918$ and $|\delta_1|\epsilon_1=0.9008\times 10^{-2}$. Then, we have $\epsilon_1=2.032,|\Delta_3|=1.557\times 10^{-2},\delta_3\epsilon_2\equiv A=3.90\times 10^{-2}$. Requiring $a_2|\Delta_1| \delta_3\epsilon_2+a_3\Delta_2|\delta_1| \epsilon_1+a_4|\Delta_2| |\Delta_3|  =2.99\times 10^{-4}$, where all term are O(10$^{-4}$).  
 To obtain the relations between phases, $\Delta_{\rm ph1}=\Delta_{\rm ph3}=\delta$, we can consider the following superpotential,
\dis{
 W_{CP}=-i\,\mu_1\delta_1\delta_3+\frac{1}{\mu_1}\delta_1^4+i\,M_0^2\Delta_1+M_1\Delta_1^2+M_2\Delta_2^2+i\,\lambda_1\Delta_1\Delta_2^2+i\,\lambda_2 \Delta_3\delta_3\epsilon_2
+\lambda_3 \Delta_2\Delta_3^2 +\frac{1}{\mu_2^2}\delta_1\Delta_3^3 \epsilon_1,
}
where parameters are real numbers.
The following SUSY conditions lead to the desired relations:
\dis{
& \frac{4}{\mu_1}\delta_1^3 -i\,\mu_1\delta_3+\frac{1}{\mu_2^2} \Delta_3^3 \epsilon_1=0 ,\\
& -\mu_1 \delta_1  +\lambda_2\Delta_3\epsilon_2=0 ,\\
& i\,M_0^2+i\,\lambda_1 \Delta_2^2+2M_1\Delta_1=0 ,\\
& i\,2\lambda_1 \Delta_1\Delta_2+\lambda_3\Delta_3^2+2 M_2\Delta_2=0 ,\\
 &i\, \lambda_2\delta_3\epsilon_2+2\lambda_3\Delta_2\Delta_3
+\frac{3}{\mu_2^2}\delta_1\Delta_3^2 \epsilon_1
 =0.\\
 }

%%%%%%%
\subsection{Lepton mass matrices}
Again, we use the KS parametrization \cite{KimSeo11} to specify the phase $\delta_L=\delta_{\rm PMNS}$ from the (3,1) element of $M_e$. Note that  the preliminary value   $\delta_{\rm PMNS}\approx -\frac{\pi}{2}$ \cite{T2K17},
 \dis{
 U_{\rm PMNS}=
\begin{pmatrix}
\quad C_1,\quad   &   +S_1C_3,\quad &\quad~~+S_1S_3\quad~~ \\[0.5em]
-C_2S_1,& +C_1C_2C_3+S_2S_3e^{-i\delta_L}, \quad &+C_1C_2S_3  -S_2C_3e^{-i\delta_L}\\[0.5em]
 -S_1S_2e^{i\delta_L}, &-C_2S_3 +C_1S_2C_3 e^{i\delta_L}, \quad & +C_2C_3+C_1S_2S_3e^{i\delta_L}
\end{pmatrix},\label{eq:PMNSpara}
}
where the parameters  are the leptonic parameters, $\Theta_{1,2,3}$ and $\delta_L$. Since the PMNS matrix elements are not known as accurately as the CKM matrix elements, we do not present a detail study of the leptonic sector. But note that the phase $\delta_L$ in Eq. (\ref{eq:PMNSpara}) is the PMNS phase $\delta_{\rm PMNS}$.
 
 %%%%%%%%%%%
\section{From $\EE8$ heterotic string}\label{sec:string}
In this section, we attempt to realize the texture of  quark mass matrix discussed in Subsec. \ref{subsec:quarks}. We will not discuss  the texture of neutrino mass matrix since the PMNS matrix elements are not known very accurately. Nevertheless, we will comment on the relation of CP phases in the quark and lepton sectors in this section.

  %%%%%%%%%%%%%%%%%
\begin{table}[b!]
\begin{center}
\begin{tabular}{@{}lccc||ccccc@{}} \toprule
SU(5)$_{\rm flip}$ & Sect. &~\Uanom&~$Q_1/2$ &\,SU(5)$_{\rm flip}$ & Sect. &~\Uanom &~$Q_1/2$ \\
(Symbol) & & ~($Q_{\rm anom} $)&  &(Symbol) & & ~($Q_{\rm anom} $)  \\ \colrule
 $\one_{-5}(S_1)$  & $U$  &    $+5$ & $-3$ &$\one_{0}(\sigma_1)$ &$T_4^0$ &    $-12$& $-4$ \\
$\one_{-5}(S_{24}^a)$  &$T_4^0$&   $-3$& $-1$ &  $\one_{0}(\sigma_2)$ &$T_4^0$ &    $-2$& $-4$ \\
$\one_{-5}(S_{24}^b)$  &  $T_4^0$ &   $-3$&  $-1$&  $\one_{0}(\sigma_3)$ & $T_4^0$ &  $-8$& $+2$  \\
$\tenb_{-1}(C_2)$&$U$  &   $-13$& $-3$ & $\one_{0}(\sigma_4)$ & $T_4^0$ &  $+10$& $+2$  \\
  $\five_{+3}(C_1)$ &$U$  &   $-1$& $+3$  & $\one_{0}(\sigma_5)$ &  $T_6$ &  $+14$& $0$\\
$\five_{+3}(C_{3a})$  &  $T_4^0$ &   $-3$& $-1$  & $\one_{0}(\sigma_6)$ &  $T_6$ &   $-4$& $0$\\
  $\five_{+3}(C_{3b})$  &$T_4^0$ &   $-3$& $-1$ &   $\one_{0}(\sigma_9)$  &  $T_2^0$ & $-6$& $-2$ \\
$\tenb_{-1}(C_{4a})$  &  $T_4^0$  &   $-3$& $-1$ &$\one_{0}(\sigma_{10})$  & $T_2^0$ &    $-6$& $-2$  \\
$\tenb_{-1}(C_{4b})$  & $T_4^0$  &   $-3$& $-1$ & $\one_{0}(\sigma_{13})$ & $T_3$ &    $+\frac{124}{7}$& $+3$  \\
$\tenb_{-1}(C_{11})$   &  $T_3$ &    $-\frac{33}{7}$& $0$ &$\ten_{+1}(C_{12})$   &  $T_9$ & $+\frac{33}{7}$& $0$  \\
 %\colrule
$\five_{-2}(H_{u})$   &$T_6$  &    $0$& $0$ &  $\one_{0}(\sigma_{15})$ & $T_9$ &   $+\frac{30}{7}$& $+3$ \\
 $\fiveb_{+2}(H_{d})$   &$T_6$  &   $0$& $0$ &    $\one_{0}(\sigma_{21})$ & $T_1^0$ &  $+\frac{12}{7}$& $-1$ \\[0.2em]
 \botrule
\end{tabular} \label{tab:ab1} 
\end{center}
\caption{The \flip~fields.  Fields needed in the SM are on the four left columns and SM singlet components needed at the GUT scale toward the FN mechanism are  on the four right columns. Both neutrino components in $C_{11}$ and $C_{12}$ develop a GUT scale VEV to break \flip\,down to the SM.}
\end{table}
 
 %%%%%
\subsection{$\Z_{12-I}$ orbifold compactification}
%%%%%%%%%%%%%%%

Note that the SM mass matrix 
\dis{
\bar{\Psi}_{I\,L}^cC^{-1}\Psi_{J\,L} M^{IJ}+{\rm h.c.}
}
gives in general non-symmetric mass matrix of $M$ because ${\Psi}_{L}^c$ and $\Psi_{L} $ transform differently under SU(2)$_L\times$U(1)$_Y$.  In the GUT model, Majorana neutrinos in the SU(2)$_L$ doublets are embedded in $\five_0$ of SU(5) in the GG model and $\five_{+3}$ in the \flip. Then, the effective neutrino mass matrix  in these simple GUTs are symmetric.  For the quark mass matrix, $\tenb_f \tenb_f\fiveb_{\rm Higgs}$  is the up-type quark mass matrix in the GG model, which is symmetric. In the GG model, we usually use diagonalized  up-type quark mass matrix, and consider non-symmetric $\five_f \tenb_f\five_{\rm Higgs}$ for the down-type quark mass matrix. On the other hand, in the \flip~ the down-type quark mass matrix, $\tenb_f \tenb_f\fiveb_{\rm Higgs}$  is symmetric  and the up-type quark mass matrix  $\five_f \tenb_f\five_{\rm Higgs}$ is non-symmetric. So, we prefer to consider  a symmetric down-type quark mass matrix in \flip.  The up-type quark mass matrix is  non-symmetric,  and we can assign different coefficients for $M_{(u)IJ}$ and $M_{(u)JI}$,
\dis{
&{\rm down~type~quark~mass~matrix=symmetric}\\
&{\rm up~type~quark~mass~matrix=asymmetric} \label{eq:Massformflip}
}

The \flip\,GUT  gauge group presented in Ref. \cite{KimKyaeNam17} is
\dis{
SU(5)\times U(1)_X\times SU(5)'\times U(1)^{6},
}
where, in the notation of \cite{LNP696},
\dis{
X= \left(-2,-2,-2,-2,-2,~0,~0,~0 \right)(0^8)'.\label{eq:X7}
}
and  six U(1) directions of Ref. \cite{KimKyaeNam17} are
\dis{
&Q_1=\left( 0^5,12,0,0\right)\left(0^8 \right)'  , \\[0.2em]
&Q_2=\left( 0^5,0,12,0\right)\left(0^8 \right)' ,  \\[0.2em]
&Q_3=\left( 0^5,0,0,12\right)\left(0^8 \right)'  , \\[0.2em]
 &Q_4=\left(0^8 \right)\left(0^4,0,12,-12,0 \right)',\\[0.2em]
 &Q_5=\left(0^8 \right)\left(0^4,0,-6,-6,12 \right)',\\[0.2em]
&Q_6=\left(0^8 \right)\left(-6,-6,-6,-6,18,0,0,6 \right)'.
  \label{eq:Qorig}
}
 $Q_{\rm anom}$ is given by
\dis{
Q_{\rm anom} &= \frac{1}{126} (84Q_1 + 147Q_2 
-42Q_3 -63Q_5 - 9Q_6).
}
We will use  notations of Ref. \cite{KimKyaeNam17}  for the names of the fields, twisted sectors ($T_1^0, \cdots, T_6$) and the untwisted sector ($U$). We also list $\frac12 Q_{1}$ in Tables IV  such that a discrete subgroup of $U(1)_1$ can be used for matter parity if needed.  We choose one gauged U(1) example beyond \Uanom, and we checked that any other choice leads to the same conclusion.
%%%%%%%%%%
\subsection{Doublet-triplet splitting}
In the \flip, it is well-known that there is a possibility of doublet-triplet splitting. $C_{12}$ and $C_{14}$ in Eq. (\ref{eq:GUTvev}) develop the GUT scale VEVs,
\dis{
&\langle C_{11}\rangle=\langle C_{12}\rangle\equiv V\\
&\langle C_{14}\rangle=-V s_\beta c_\gamma
}
where the first equality is for vanishing D-term at the GUT scale. The renormalizable coupling, including the Higgs quintet $ \ten_{+1}\five_{-2}\ten_{+1}\sim \Phi^{ab}\Phi^c\Phi^{de}\epsilon_{abcde}$ might give  the GUT scale mass term to colored scalars by $\{de\}=\{45\}$, but  $C_{12}H_u C_{12}$ coupling is not allowed by the non-vanishing $Q_{\rm anom}$. A possible higher dimensional operator consistent with the orbifold selection rules and \Uanom$\times$U(1$)_3$ gauge symmetry is 
\dis{
\frac{1}{M^4}C_{11}[\tenb_{-1}(T_3)]H_d [\fiveb_{+2}(T_6)]C_{11}[\tenb_{-1}(T_3)]\sigma_3[\one(T_4^0)]\sigma_5[\one(T_6)]\sigma_{21}[\one(T_1^0)]\sigma_{21}[\one(T_1^0)]
}
By giving GUT to Planck scale VEVs to $C_{11},\sigma_{3},\sigma_{5}$, and $\sigma_{21}$, we obtain a GUT scale mass term for colored scalars,
\dis{
M_T\epsilon_{\alpha\beta\gamma}\Phi^{\alpha\beta}\Phi^\gamma
}
where $\alpha,\beta,\gamma$ are the color indices. Thus, the color anti-triplet in $\tenb$ combines with the color triplet in the Higgs quintet $\fiveb$. The colored scalar in the Higgs quintet $H_d$ is removed at the GUT scale, and there remains just the Higgs doublet from $H_d$. For this doublet-triplet splitting, we need
\dis{
\langle\sigma_{3}\rangle\ne 0,~
\langle\sigma_{5}\rangle\ne 0,~
\langle\sigma_{21}\rangle\ne 0,\label{eq:Svevs1}
}
and the color triplet mass is estimated as
\dis{
M_T\simeq V\frac{\langle\sigma_{3}\rangle\langle\sigma_{5}\rangle \langle\sigma_{21}\rangle^2 }{M^4}.\label{eq:MTest}
}
Suppose $V\sim  M,\sigma_{3,5}\sim 10^{-2.5} M$ and $\langle\sigma_{21}\rangle^2\sim 10^{-1} M$.  Then, we obtain $M_T\sim 10^{-6}M\sim 0.6\times 10^{12}\gev$ for $M\sim 6\times 10^{17}\gev$. $10^{12}\gev$ colored scalar with small Yukawa couplings of the first family is acceptable.
Similarly, considering  $C_{12}H_u C_{12}\sim \ten_{+1}\five_{-2}\ten_{+1}$,
\dis{
\frac{1}{M^4}C_{12}[\ten_{+1}(T_9)]H_u [\five_{-2}(T_6)]C_{12}[\ten_{+1}(T_9)]\sigma_1[\one(T_4^0)]\sigma_9[\one(T_2^0)]\sigma_{15}[\one(T_9)]\sigma_{15}[\one(T_9)]
}
the colored scalar in the Higgs quintet $H_u$ is removed at the GUT scale and there remains just the Higgs doublet from $H_u$.
 For this, we further require
\dis{
\langle\sigma_{1}\rangle\ne 0,~
\langle\sigma_{9}\rangle\ne 0,~
\langle\sigma_{15}\rangle\ne 0.\label{eq:Svevs2}
}

%%%%%%%%%%
\subsection{Proton decay problem}\label{subsec:pdecay}

%%%%%%%%%%%%%%%%%%
\begin{figure}[!t]
\includegraphics[width=0.39\textwidth]{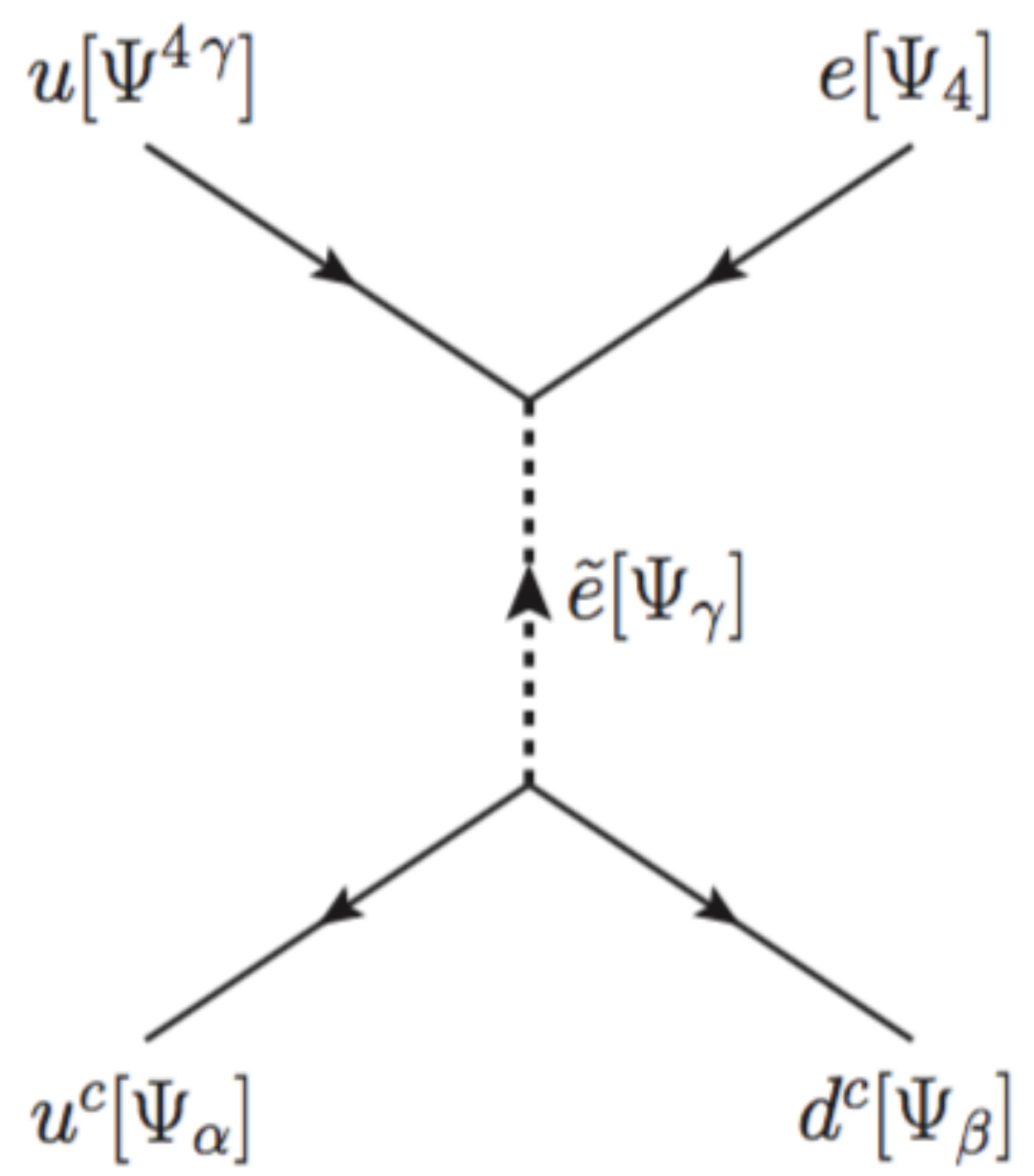} 
  \caption{A diagram for $\Delta B\ne 0$.} \label{fig:pDecay} 
\end{figure}
%%%%%%%%%%%%%%%%
 
 One may consider another gauge symmetry to obtain a $\Z_2$ discrete group by breaking U(1)$_1$ by some VEVs of singlet fields carrying even  quantum numbers of $Q_1/2$ in Table IV. It can serve as a kind of matter parity since \flip~matter fields carry odd  $Q_1/2$. But this discrete symmetry does not work because $\langle \sigma_{21}\rangle$ of Eq. (\ref{eq:Svevs1}) and $\langle \sigma_{15}\rangle$ of Eq. (\ref{eq:Svevs2})   carry the odd quantum number of  $Q_1/2$.  We do not have any mechanism for matter parity. The proton decay amplitude must be estimated in detail.\footnote{If an R parity is introduced \cite{Lee11}, the proton decay problem is automatically solved.}
 
In SUSY models, the dimension 5 proton decay operator must be sufficiently suppressed \cite{Dermisek00}.
The dimension 5 proton decay operators to electronic and muonic leptons are from the superpotential $q^1q^1q^1l^{1,2}$, \ie  $C_{15}C_{15}C_{15}C_{17}$ and $C_{15}C_{15}C_{15}C_{16}$.  Note that $C_{15},C_{17}$, and $C_{16}$ are allowed from the sector $T_4^0$. Therefore, the $\Z_{12-I}$ orbifold selection rules forbid the  product of these four fields from  $T_4^0$, and hence   there is no serious proton decay problem from the above dimension 5 operator multiplied by FN singlets ($\sigma$'s)  appear at least at dimension 7 level in our $\Z_{12-I}$ model.
 
 If it were the GG SU(5),  the cubic superpotential written in terms of matter parity violating term, $\tenb_0\five_0\five_0$, triggers proton decay as shown in Fig. \ref{fig:pDecay}  \cite{RabyBk}.  In the \flip~also there arise dangerous proton decay operators
 \dis{
 \tenb_{-1}^m\tenb_{-1}^m\five_{+3}^m\tenb_{-1}^H,~ \five_{+3}^m\five_{+3}^m\one^m_{-5}\tenb_{-1}^H, \label{eq:pDecay}
 }
 where fields with superscript $m$ are matter fields and $\tenb_{-1}^H$ is the field breaking \flip~to the SM. The above operators trigger proton decay in our model by products of FN singlets ($\sigma$'s) appear at dimension 10 level,
 \dis{
 &C_4[\tenb_{-1}(T_4^0)]C_4[\tenb_{-1}(T_4^0)]C_3[\five_{+3}(T_4^0)]\frac{1}{M^6}\left\{C_{11}[\tenb_{-1}(T_3)]\sigma_3[\one(T_4^0)]\sigma_3[\one(T_4^0)]\sigma_5[\one(T_6)]\sigma_5[\one(T_6)] \sigma_{21}[\one(T_1^0)]\right\},\\
&  C_3[\five_{+3}(T_4^0)]C_3[\five_{+3}(T_4^0)]S_{24}[\one_{-5}(T_4^0)] \frac{1}{M^6}\left\{ C_{11}[\tenb_{-1}(T_3)]\sigma_3[\one(T_4^0)]\sigma_3[\one(T_4^0)]\sigma_5[\one(T_6)]\sigma_5[\one(T_6)] \sigma_{21}[\one(T_1^0)]\right\}.\label{eq:pDecRen2}
 }
All the singlets appearing in  Eq. (\ref{eq:pDecRen2}) are the needed fields for the doublet-triplet splitting in Eq. (\ref{eq:Svevs1}). The coupling  in Fig.  \ref{fig:pDecay} is estimated, from the first term of Eq. (\ref{eq:pDecRen2}) for example as,
\dis{
\left\{\frac{1}{M^6}C_{11} \sigma_3^2\sigma_5^2\sigma_{21}\right\}^2 \left(M_{\rm SUSY}\right)^{-2}\sim\frac{1}{\Mg^2}   \frac{C_{11}^2\sigma_3^4\sigma_5^4 \sigma_{21}^2}{M^{12} }   \left(\frac{\Mg}{ M_{\rm SUSY} }\right)^2.
}
Suppose that the SUSY breaking scale $M_{\rm SUSY}\sim 10\tev$, the GUT scale $\Mg\approx 3\times 10^{16}\gev$, and the compactification scale $M\sim 6\times 10^{17}\gev$. Then, the last factor $\sim  (3\times 10^{12})^2$ is balanced by $ M_{\rm vev} \lesssim 0.5\times 10^{16}\gev$ where $M_{\rm vev}$ is some average VEV of $C_{11}$ and neutral $\sigma$ fields. The estimate given in Eq. (\ref{eq:MTest})  can be fitted to this average.
Thus, the dimension 6 operator of Fig.   \ref{fig:pDecay} can be controlled such that it is not so strong as the dimension 6 operator derived from the exchange of leptoquark gauge bosons in SUSY GUTs.

%%%%%%%%%%
\subsection{Families}\label{subsec:family}

There are three $\five_{+3}$'s   and  three $\one_{-5}$'s in Table IV. These include all members of three SM lepton doublets three $u^c$-type quarks.
However, there are four $\tenb_{-1}$'s in Table IV. So, there are a few possibilities of choosing three SM quark doublets. Out of  four $\tenb_{-1}$'s, we always choose $\tenb_{-1}$ in the $U$ sector. Then, there are three possibilites of choosing two remaining quark doublets:  (1) the antisymmetric combination of   $\tenb_{-1}$'s from the $T_4^0$ sector and  $\tenb_{-1}$ from the $T_3$ sector, (2) two   $\tenb_{-1}$'s from the $T_4^0$ sector, and (3) a linear  combination of  $\tenb_{-1}$ of $T_3$ and  antisymmetric  $\tenb_{-1}$  from $T_4^0$, and a linear  combination of  $\tenb_{-1}$ of $T_3$ and  symmetric  $\tenb_{-1}$  from $T_4^0$.
 All these are considered by mixing three $\tenb_{-1}$'s, introducing three angles $\alpha,\beta$ and $\gamma$,
\dis{
&C_{13}[\tenb_{-1}(T_4^0),\tenb_{-1}(T_3)]\equiv  + C_{11}c_\beta+ C_{4a} c_\alpha s_\beta
- C_{4b} s_\alpha s_\beta,\\[0.2em]
&C_{14}[\tenb_{-1}(T_4^0),\tenb_{-1}(T_3)]\equiv  - C_{11}s_\beta c_\gamma +C_{4a}( c_\alpha c_\beta c_\gamma  - s_\alpha s_\gamma )-C_{4b}(s_\alpha c_\beta c_\gamma+ c_\alpha s_\gamma) ,\\[0.2em]
&C_{15}[\tenb_{-1}(T_4^0),\tenb_{-1}(T_3)]\equiv  - C_{11}s_\beta s_\gamma +C_{4a}( s_\alpha  c_\gamma + c_\alpha c_\beta s_\gamma )+C_{4b}( c_\alpha   c_\gamma -s_\alpha c_\beta s_\gamma).\label{eq:tenbs}
}
where $s_{\alpha,\beta,\gamma}=\sin_{\alpha,\beta,\gamma}$ and  $c_{\alpha,\beta,\gamma}=\cos_{\alpha,\beta,\gamma}$. We choose two out of the above three combinations.
Similarly, we define
\dis{
&C_{16}[\five_{+3}(T_4^0)]\equiv \frac{1}{\sqrt2}\Big(+C_{3a}+C_{3b}\Big),\\[0.5em]
&C_{17}[\five_{+3}(T_4^0)]\equiv \frac{1}{\sqrt2}\Big(-C_{3a}+C_{3b}\Big).\label{eq:fives}
}
Now, let us identify $\ten_{+1}$ and $\tenb_{-1}$'s of Table I   as
\dis{
C_{12}[\ten_{+1}]\oplus C_{14}[\tenb_{-1}]:~{\rm The~Higgs~set~for~breaking~}SU(5)\times U(1)_X,\label{eq:GUTvev}
}
and
\dis{
&C_{15} :~{\rm 1st~family},\\[0.3em]
&C_{13} :~{\rm 2nd~family},\\[0.3em]
&C_2 :~{\rm 3rd~family},\label{eq:Modelu}
}  
and $\five_{+3}$'s of Table I   as
\dis{
&C_{17} :~{\rm 1st~family},\\[0.3em]
&C_{16} :~{\rm 2nd~family},\\[0.3em]
&C_1 :~{\rm 3rd~family},\label{eq:Modeld}
}  

%%%%%%%%%
\vskip 0.5cm
In this paper, it is outside the scope of current analysis to see the details of superpotential.  We just assume ceratin VEVs to fit to the observed data.

%%%%%%%%%%%%%
\subsubsection{Down-type quarks}
Let us scale scalar fields and mass matrices such that they are made dimensionless by dividing with a mass parameter, for example by $M$.

The down-type quark masses are
 \dis{
&M_{d\,(11)}^w =C_{15}[\tenb_{T_3}]C_{15}[\tenb_{T_3}] H_d(\fiveb_{T_6})\,\sigma_1(\one_{T_4^0})\sigma_6(\one_{T_6})\sigma_6(\one_{T_6}) \sigma_{9}(\one_{T_2^0})\sigma_{13}(\one_{T_3})\sigma_{13}(\one_{T_3}),
\\[0.2em]
&M_{d\,(22)}^w = C_{13}(\tenb_{T_4^0})C_{13}(\tenb_{T_4^0}) H_d(\fiveb_{T_6})\Big\{\sigma_4(\one_{T_4^0})\sigma_6(\one_{T_6}) , \sigma_{15}(\one_{T_9})\sigma_{21}(\one_{T_1^0})\Big\},\\[0.2em]
&M_{d\,(33)}^w=C_2(\tenb_{U})C_2(\tenb_{U})H_d(\fiveb_{T_6})\sigma_5(\one_{T_6})\sigma_3(\one_{T_4^0})\sigma_4(\one_{T_4^0})\sigma_4(\one_{T_4^0}), \label{eq:downdiag}
}
where we presesented only the antisymmetric part in $M_{d\,(22)}^w$ and  only the component from $T_3$ in $M_{d\,(11)}^w $. For the down-type quarks, it is enough to show non-zero $M_{d\,(33)}^w$ and $M_{d\,(22)}^w$  and the  conditions for making the off-diagonsal elements vanish,
 \dis{
&M_{d\,(12)}^w=C_{15}(\tenb_{T_4^0})C_{13}[\tenb_{T_3}]H_d(\fiveb_{T_6}))\Big\{ \sigma_2(\one_{T_4^0})\sigma_3(\one_{T_4^0})\sigma_{13}(\one_{T_3}),\sigma_6(\one_{T_6})\sigma_9(\one_{T_2^0})\sigma_{13}(\one_{T_3})\Big\} =0 ,\\[0.2em]
&M_{d\,(21)}^w=C_{13}[\tenb_{T_3}]C_{15}(\tenb_{T_4^0}) H_d(\fiveb_{T_6})\Big\{  \sigma_2(\one_{T_4^0})\sigma_3(\one_{T_4^0})\sigma_{13}(\one_{T_3}) , \sigma_6(\one_{T_6}) \sigma_{9}(\one_{T_2^0})\sigma_{13}(\one_{T_3})\Big\}=0,\\[0.2em]
&M_{d\,(13)}^w=C_{15}(\tenb_{T_4^0})C_2(\tenb_{U}) H_d(\fiveb_{T_6})\Big\{c\,\sigma_4(\one_{T_4^0})\sigma_4(\one_{T_4^0})\sigma_6(\one_{T_6})+c'\sigma_4(\one_{T_4^0}) \sigma_{15}(\one_{T_9})\sigma_{21}(\one_{T_1^0})\Big\}=0,\\[0.2em]
&M_{d\,(31)}^w=C_2(\tenb_{U})C_{15}(\tenb_{T_4^0}) H_d(\fiveb_{T_6})\Big\{c\,\sigma_4(\one_{T_4^0})\sigma_4(\one_{T_4^0})\sigma_6(\one_{T_6})+c'\sigma_4(\one_{T_4^0}) \sigma_{15}(\one_{T_9})\sigma_{21}(\one_{T_1^0}\Big\}=0,\\[0.2em]
&M_{d\,(23)}^w=C_{13}[\tenb_{T_3}]C_2(\tenb_{U}) H_d(\fiveb_{T_6})   \sigma_{13}(\one_{T_3})=0,\\[0.2em]
&M_{d\,(32)}^w=C_2(\tenb_{U})C_{13}[\tenb_{T_3}]H_d(\fiveb_{T_6})\sigma_{13}(\one_{T_3})=0.
\label{eq:downoff}
}
To satisfy the conditions of Eq. (\ref{eq:downoff}), let us choose
\dis{
\langle  \sigma_{13}\rangle  =0, \label{eq:VEV13}
}
and
\dis{
c\langle  \sigma_4 \sigma_6\rangle  +c'\langle  \sigma_{15}\sigma_{21}\rangle  =0.\label{eq:ratiosfit}
}
$M_{d\,(13)}^w$ and $M_{d\,(31)}^w$ can be made to vanish.

%%%%%%%%%%%%%%
\subsubsection{Up-type quarks}
 
 Therefore, we consider the $W^-_\mu$ coupling instead of $W^+_\mu$  coupling of Eq. (\ref{eq:Wplus}), $V_u^\dagger M^{\rm diag}_u U_u$ as 
\dis{
\begin{pmatrix}
\begin{array}{l} c_4c_1 m_u+c_5s_4c_2s_1m_c\\
+s_4s_5s_1s_2e^{-i\Delta+i\delta} m_t\\ ,\end{array}   &\begin{array}{l}  c_4s_1c_3m_u -c_5s_4c_1c_2c_3m_c\\
-c_5s_4s_2s_3e^{-i\delta}m_c+s_4s_5c_2s_3e^{-i\Delta} m_t\\
-s_4s_5c_1s_2c_3e^{-i\Delta+i\delta} m_t,
\end{array} &\begin{array}{l}c_4s_1s_3 m_u   -c_5s_4c_1c_2s_3m_c\\ +c_5s_4s_2c_3e^{-i\delta}m_c  
-s_4s_5c_2c_3e^{-i\Delta} m_t\\
-s_4s_5c_1s_2s_3e^{-i\Delta+i\delta} m_t,\end{array}  \\[1em]
&&\\
\begin{array}{l} s_4c_6c_1 m_u-c_4c_5c_6 c_2s_1m_c\\
-s_5s_6c_2s_1 e^{i\Delta}m_c\\ 
+c_5s_6s_1s_2 e^{i\delta}m_t \\
-c_4s_5c_6 s_1s_2e^{-i\Delta+i\delta}m_t\\ \\ ,\end{array}  & 
\begin{array}{l} s_4c_6s_1c_3 m_u+c_4c_5c_6 c_1c_2c_3m_c\\
+c_4c_5c_6 s_2s_3e^{-i\delta}m_c\\
+s_5s_6 c_1c_2c_3e^{i\Delta}m_c+s_5s_6 s_2s_3e^{i\Delta-i\delta}m_c\\ 
+c_5s_6c_2s_3 m_t - c_5s_6c_1s_2c_3e^{i\delta}m_t\\
-c_4s_5c_6 c_2s_3e^{-i\Delta}m_t\\
+c_4s_5c_6 c_1s_2c_3e^{-i\Delta+i\delta}m_t,\end{array}  
& 
\begin{array}{l} s_4c_6s_1s_3 m_u+c_4c_5c_6 c_1c_2s_3m_c\\
-c_4c_5c_6 s_2c_3e^{-i\delta}m_c\\
+s_5s_6 c_1c_2s_3e^{i\Delta}m_c-s_5s_6 s_2c_3e^{i\Delta-i\delta}m_c\\ 
-c_5s_6c_2c_3 m_t - c_5s_6c_1s_2s_3e^{i\delta}m_t\\
+c_4s_5c_6 c_2c_3e^{-i\Delta}m_t\\
+c_4s_5c_6 c_1s_2s_3e^{-i\Delta+i\delta}m_t,\end{array}  \\[1em]
&&\\
 \begin{array}{l} s_4s_6c_1 m_u-c_4c_5s_6 c_2s_1m_c\\
+s_5c_6c_2s_1 e^{i\Delta}m_c\\ 
-c_5c_6s_1s_2 e^{i\delta}m_t \\
-c_4s_5s_6 s_1s_2e^{-i\Delta+i\delta}m_t\\ \\ ,\end{array}   
&\begin{array}{l} s_4s_6s_1c_3 m_u+c_4c_5s_6 c_1c_2c_3m_c\\
+c_4c_5s_6 s_2s_3e^{-i\delta}m_c\\
-s_5c_6 c_1c_2c_3e^{i\Delta}m_c-s_5c_6 s_2s_3e^{i\Delta-i\delta}m_c\\ 
-c_5c_6c_2s_3 m_t + c_5c_6c_1s_2c_3e^{i\delta}m_t\\
-c_4s_5s_6 c_2s_3e^{-i\Delta}m_t\\
+c_4s_5s_6 c_1s_2c_3e^{-i\Delta+i\delta}m_t,\end{array}  
& \begin{array}{l} s_4s_6s_1s_3 m_u+c_4c_5s_6 c_1c_2s_3m_c\\
-c_4c_5s_6 s_2c_3e^{-i\delta}m_c\\
-s_5c_6 c_1c_2s_3e^{i\Delta}m_c+s_5c_6 s_2c_3e^{i\Delta-i\delta}m_c \\ 
+c_5c_6c_2c_3 m_t+ c_5c_6c_1s_2s_3e^{i\delta}m_t\\
+c_4s_5s_6 c_2c_3e^{-i\Delta}m_t\\
+c_4s_5s_6 c_1s_2s_3e^{-i\Delta+i\delta}m_t,\end{array} 
\end{pmatrix}\label{eq:Wminus}
}
Change the sign $m_u\to -m_u$, and to reduce the number of parameters let us  choose parameters of R-fields as
\dis{
\frac{s_4}{c_4}=\frac{s_1}{c_1} , ~\frac{s_5}{c_5}=\frac{m_c}{m_t}\frac{s_2}{c_2},~ s_6=0,~ \Delta=\delta.  
}
Then, we obtain  
\dis{
V_u^\dagger M^{\rm diag}_u U_u = \begin{pmatrix}
  -c_4c_1 m_u+c_5s_4c_2^{-1}s_1 m_c ,  &   -c_4s_1c_3m_u -c_5s_4c_1c_2^{-1}c_3m_c,
&  -c_4s_1s_3 m_u   -c_5c_4s_1c_2^{-1}s_3m_c \\[0.3em]
\\
\begin{array}{l} -s_4c_1 m_u \\
-c_4c_5  s_1 c_2^{-1} m_c ,\end{array}  & 
\begin{array}{l} -s_4 s_1c_3 m_u 
\\+c_4c_5  c_1c_3 c_2^{-1} m_c,\end{array}  
& 
\begin{array}{l} -s_4 s_1s_3 m_u
\\
+c_4c_5  c_1 s_3 c_2^{-1} m_c,\end{array}  \\[0.3em]
\\
 \begin{array}{l}  
+s_5 c_2s_1 e^{i\delta}m_c\\ 
-c_5 s_1s_2 e^{i\delta}m_t  ,\end{array}   
&\begin{array}{l}   
-s_5  c_1c_2c_3e^{i\delta}m_c-s_5  s_2s_3 m_c \\ 
-c_5 c_2s_3 m_t + c_5 c_1s_2c_3e^{i\delta}m_t ,\end{array}  
& \begin{array}{l}   
-s_5  c_1c_2s_3e^{i\delta}m_c+s_5  s_2c_3 m_c \\ 
+c_5 c_2c_3 m_t+ c_5 c_1s_2s_3e^{i\delta}m_t
 ,\end{array} 
\end{pmatrix}
}
where we require $c_2,c_3,c_5\simeq$ O(1). Also, $s_5$ can be O(1). Thus, we consider,
\dis{
V_u^\dagger M^{\rm diag}_u U_u =m_t\begin{pmatrix}
  -c_4 c_1   \frac{m_u}{c_5m_t}+ s_4 s_1  \frac{m_c}{m_t} ,  &   -c_4s_1  \frac{m_u}{c_5m_t} - s_4c_1  \frac{m_c}{m_t},
&   - c_4s_1 s_3 \frac{m_c}{m_t}  \\[1em]
  -s_4 c_1   \frac{m_u}{c_5m_t} 
-c_4  s_1   \frac{m_c}{m_t}  , & +c_4  c_1  \frac{m_c}{m_t}, 
&  
+c_4 c_1 s_3  \frac{m_c}{m_t}\\[1em]
- s_1( s_2 - s_5c_2 \frac{m_c}{c_5m_t} ) e^{i\delta}\frac{m_c}{m_t}  , 
&\Big[-  s_3   + c_1(  s_2-s_5c_3\frac{m_c}{c_5m_t} ) e^{i\delta}\Big]\frac{m_c}{m_t}   ,&  1
\end{pmatrix}
}
where we neglected $m_t s_2s_3, m_cs_2, m_cs_3$.
\dis{
&s_1=0.2252, c_1=0.9743, s_2=0.0400, c_2=0.9992, s_3=0.01557, c_3=0.9999,\\
&m_u=2.5\,{\rm MeV}, m_c=1280\,{\rm MeV}, m_t=173\,{\rm GeV}, \frac{m_u}{m_t}=1.45\times 10^{-5}  ,\frac{m_c}{m_t}  =0.74\times 10^{-2},
}
so that $M_u^w/m_t$ is approximately given by
 \dis{
 \begin{pmatrix}
 + 1.67\times 10^{-3}s_4 ,  &    - 0.721\times 10^{-2}s_4,
&   - 2.59\times 10^{-5}c_4 \\[1em]
  -1.67\times 10^{-3}c_4  , & + 0.7 21\times 10^{-2}c_4 , 
&  
+1.12\times 10^{-4}c_4  \\[1em]
 (-0.67\times 10^{-6} +1.24\times 10^{-5}t_5  ) e^{i\delta}   , 
& -  1.15\times 10^{-3} +(2.88\times 10^{-4} -0.534\times 10^{-4}t_5  ) e^{i\delta}  ,&  1
\end{pmatrix}\label{eq:MupFit}
}

The up-type quark masses are
 \begin{eqnarray}
&M_{u\,(11)}^w&=0,\nonumber \\[0.2em]
&M_{u\,(22)}^w&=C_{16}(\five_{T_4^0})C_{13}(\tenb_{T_4^0},\tenb_{T_3}) H_u(\five_{T_6}) \Big\{ 
\sigma_{4}(T_4^0)\sigma_{6}(T_6), \sigma_{15}(T_9)\sigma_{21}(T_1^0); \sigma_{4}(T_4^0) \sigma_{6}(T_6)\sigma_{21}(T_1^0)   \Big\},\nonumber \\[0.2em]
&M_{u\,(33)}^w&=C_1(\five_{U})C_{2}(\tenb_{U}) H_u(\five_{T_6}) \sigma_5(\one_{T_6}),\nonumber
\end{eqnarray}

\begin{eqnarray}
&M_{u\,(12)}^w&=C_{17}(\five_{T_4^0})C_{13}(\tenb_{T_4^0},\tenb_{T_3}) H_u(\five_{T_6})\Big\{\sigma_{3}(T_4^0)\sigma_{5}(T_6) , \sigma_{15}(T_9)\sigma_{21}(T_1^0) ;\nonumber\\
 &&\qquad     \sigma_2(\one_{T_4^0})\sigma_3(\one_{T_4^0})\sigma_{13}(\one_{T_3}), \sigma_6(\one_{T_6})\sigma_9(\one_{T_2^0})\sigma_{13}(\one_{T_3}), \sigma_{21}(T_1^0)\sigma_{3}(T_4^0)\sigma_{5}(T_6)
\Big\},\nonumber\\[0.2em]
&M_{u\,(21)}^w&=C_{16}(\five_{T_4^0})C_{15}(\tenb_{T_4^0},\tenb_{T_3}) H_u(\five_{T_6})\Big\{ \sigma_{3}(T_4^0)\sigma_{5}(T_6) ,\sigma_{15}(T_9)\sigma_{21}(T_1^0) ;\nonumber\\
 &&\qquad     \sigma_2(\one_{T_4^0})\sigma_3(\one_{T_4^0})\sigma_{13}(\one_{T_3}), \sigma_6(\one_{T_6})\sigma_9(\one_{T_2^0})\sigma_{13}(\one_{T_3}), \sigma_{21}(T_1^0)\sigma_{3}(T_4^0)\sigma_{5}(T_6)
\Big\},\nonumber\\[0.2em]
  &M_{u\,(23)}^w&=C_{16}(\five_{T_4^0})C_{2}(\tenb_{U}) H_u(\five_{T_6})\sigma_3(\one_{T_4^0})\sigma_4(\one_{T_4^0})\sigma_5(\one_{T_6}),\nonumber
\\[0.2em]
  &M_{u\,(32)}^w&=C_1(\five_{U})C_{13}(\tenb_{T_4^0},\tenb_{T_3})  H_u(\five_{T_6})\Big\{
 \sigma_{5}(\one_{T_6})\sigma_6(\one_{T_6})\sigma_{9}(\one_{T_2^0}),  \sigma_2(\one_{T_4^0}) \sigma_4(\one_{T_4^0})\sigma_{6}(\one_{T_6});\nonumber \\
&&\qquad    \sigma_{13}(\one_{T_3}) \sigma_{2}(\one_{T_4^0})\sigma_6(\one_{T_6})\sigma_{9}(\one_{T_2^0}),   \sigma_{13}(\one_{T_3}) \sigma_{2}(\one_{T_4^0})^2\sigma_3(\one_{T_4^0}),  \ \sigma_{21}(\one_{T_1^0}) \sigma_{2}(\one_{T_4^0}) \sigma_3(\one_{T_4^0}) \sigma_5(\one_{T_6});\nonumber \\
&&\qquad    \sigma_{21}(\one_{T_1^0}) \sigma_{5}(\one_{T_6}) \sigma_6(\one_{T_6}) \sigma_9(\one_{T_2^0}),  \sigma_{2}(\one_{T_4^0}) \sigma_{15}(\one_{T_9}) \sigma_{21}(\one_{T_1^0})^2   \Big\}, \nonumber
\end{eqnarray}

\begin{eqnarray}
& M_{u\,(13)}^w&=0,\label{eq:up13}
 \\[0.2em]
 &M_{u\,(31)}^w&=C_1(\five_{U})C_{13}(\tenb_{T_4^0},\tenb_{T_3})  H_u(\five_{T_6})\Big\{
 \sigma_{5}(\one_{T_6})\sigma_6(\one_{T_6})\sigma_{9}(\one_{T_2^0}),  \sigma_2(\one_{T_4^0}) \sigma_4(\one_{T_4^0})\sigma_{6}(\one_{T_6});  \nonumber\\
&&\qquad    \sigma_{13}(\one_{T_3}) \sigma_{2}(\one_{T_4^0})\sigma_6(\one_{T_6})\sigma_{9}(\one_{T_2^0}),   \sigma_{13}(\one_{T_3}) \sigma_{2}(\one_{T_4^0})^2\sigma_3(\one_{T_4^0}),  \ \sigma_{21}(\one_{T_1^0}) \sigma_{2}(\one_{T_4^0}) \sigma_3(\one_{T_4^0}) \sigma_5(\one_{T_6});\nonumber \\
&&\qquad    \sigma_{21}(\one_{T_1^0}) \sigma_{5}(\one_{T_6}) \sigma_6(\one_{T_6}) \sigma_9(\one_{T_2^0}),  \sigma_{2}(\one_{T_4^0}) \sigma_{15}(\one_{T_9}) \sigma_{21}(\one_{T_1^0}) ^2  \Big\}.\label{eq:up31}
\end{eqnarray}
 $M_{u\,(33)}^w$ is the largest value, and we set $\langle \sigma_5\rangle=O(1)$, and automatically we have $M_{u\,(11)}^w=M_{u\,(13)}^w=0$ by the unavoidable antisymmetric property among $\five_{-3}(T_4^0)$, viz. $C_{17}$ in Eq. (\ref{eq:fives}).
 
 The following example is just showing a possibility. We have chosen $\sigma_{13}=0$  in Eq. (\ref{eq:VEV13}) to make down-type quark masses diagonal.  Let us further simplify by setting $\langle \sigma_2\rangle= 0$,
 \begin{eqnarray}
&M_{u\,(11)}^w&=0,\nonumber\\[0.2em]
&M_{u\,(22)}^w&=C_{16}(\five_{T_4^0})C_{13}(\tenb_{T_4^0},\tenb_{T_3}) H_u(\five_{T_6}) \Big\{\frac{1}{\sigma_{5}(T_6)}\sigma_{4}(T_4^0)\sigma_{6}(T_6),\frac{\sigma_{15}(T_9)}{\sigma_{5}(T_6)}\sigma_{21}(T_1^0);\nonumber\\
&&\qquad \qquad \frac{\sigma_{15}(T_9)}{\sigma_{5}(T_6)}\sigma_{21}(T_1^0)^2,\sigma_{3}(T_4^0)\sigma_{21}(T_1^0) \Big\}\sigma_{5}(T_6),\nonumber\\[0.2em]
&M_{u\,(33)}^w&=C_1(\five_{U})C_{2}(\tenb_{U}) H_u(\five_{T_6})\, \sigma_5(\one_{T_6}) \label{eq:mtop}
\end{eqnarray}

\dis{
M_{u\,(12)}^w&=C_{17}(\five_{T_4^0})C_{13}(\tenb_{T_4^0},\tenb_{T_3}) H_u(\five_{T_6})\Big\{ \sigma_{3}(T_4^0) ,\frac{1}{\sigma_{5}(T_6)}\sigma_{15}(T_9)\sigma_{21}(T_1^0);0\Big\}\sigma_{5}(T_6),\\[0.2em]
M_{u\,(21)}^w&=C_{16}(\five_{T_4^0})C_{15}(\tenb_{T_4^0},\tenb_{T_3}) H_u(\five_{T_6})\Big\{ \sigma_{3}(T_4^0),\frac{1}{\sigma_{5}(T_6)},\sigma_{15}(T_9)\sigma_{21}(T_1^0); \sigma_{21}(T_1^0)\sigma_{3}(T_4^0)\Big\}\sigma_{5}(T_6),\\[0.2em]
  M_{u\,(23)}^w&=C_{16}(\five_{T_4^0})C_{2}(\tenb_{U}) H_u(\five_{T_6})\left\{\sigma_3(\one_{T_4^0})\sigma_4(\one_{T_4^0})\right\}\sigma_5(\one_{T_6}),
\\[0.2em]
 M_{u\,(32)}^w&=C_1(\five_{U})C_{13}(\tenb_{T_4^0},\tenb_{T_3})  H_u(\five_{T_6})\Big\{ 0 ;\sigma_6(\one_{T_6}) \sigma_9(\one_{T_2^0})  \sigma_{21}(\one_{T_1^0})  \Big\}\sigma_{5}(\one_{T_6}),\\[0.2em]
 M_{u\,(13)}^w&=0,
 \\[0.2em]
 M_{u\,(31)}^w&=C_1(\five_{U})C_{15}(\tenb_{T_4^0},\tenb_{T_3}) H_u(\five_{T_6})\Big\{
    \sigma_6(\one_{T_6})\sigma_9(\one_{T_2^0})   ;  
\sigma_6(\one_{T_6}) \sigma_9(\one_{T_2^0})  \sigma_{21}(\one_{T_1^0})   \Big\} \sigma_{5}(\one_{T_6}). \label{eq:Mu23}
}
where the antisymmetric combination of $\tenb_{-1}$'s from $T_4^0$ is written before the semi-colon and the symmetric  combinations of $\tenb_{-1}$'s from $T_4^0$ is written after the semi-colon. Zeros indicate this symmetry properties. 
  
\dis{
 \begin{pmatrix}
 0 ,  &     (p\sigma_{3}+q \frac{\sigma_{15}\sigma_{21}}{\sigma_5})  ,
&  0 \\[1em]
 (c \sigma_{3}+d  \frac{\sigma_{15}\sigma_{21}}{\sigma_5})+ e \sigma_{3}\sigma_{21}, & f\frac{\sigma_{4}\sigma_{6}}{\sigma_5}+g\frac{\sigma_{15}\sigma_{21}}{\sigma_5}+h\frac{\sigma_{15}\sigma_{21}^2}{\sigma_5}+ k\sigma_{3}\sigma_{21}) , & \ell \sigma_3\sigma_4  \\[1em]
   a \sigma_6\sigma_9(1  +r \sigma_{21} ) , 
& b \sigma_6\sigma_9\sigma_{21} ,&  1
\end{pmatrix}
}
To present a simple numerics,  let us neglect the $\sigma_{15} $ terms. So, consider
 \dis{
 \begin{pmatrix}
 0 ,  &     p\sigma_{3}  ,
&  0 \\[1em]
 c \sigma_{3} + e \sigma_{3}\sigma_{21}, & f\frac{\sigma_{4}\sigma_{6}}{\sigma_5}+ k\sigma_{3}\sigma_{21}, & \ell \sigma_3\sigma_4  \\[1em]
   a \sigma_6\sigma_9(1  +r \sigma_{21} ) , 
& b \sigma_6\sigma_9\sigma_{21} ,&  1
\end{pmatrix}
}
Assuming hierarchies of VEVs with O(1) coefficients,
\dis{
&\sigma_4\ll 1,\\
 &-p\sigma_3=f\frac{\sigma_{4}\sigma_{6}}{\sigma_5} + k\sigma_{3}\sigma_{21}\simeq O(10^{-3}) ,\\
&c\sigma_3+e\sigma_3\sigma_{21}\simeq O(10^{-3}),\\
&a\sigma_6\sigma_9 \simeq O(10^{-5}),\\
 &\sigma_{21}\simeq O(\frac{a}{b})\ll 1,\\
 & r\sigma_{21} \simeq O(1),\\
& p\sim c\sim e\sim f\sim k\sim \ell \sim r\simeq O(1),\label{eq:VEVapprox}
}
we estimate
\dis{
 \approx \begin{pmatrix}
 0 ,  &     p\sigma_{3}  ,
&  0 \\[1em]
 c \sigma_{3} , & f\frac{\sigma_{4}\sigma_{6}}{\sigma_5} , & \ell \sigma_3\sigma_4  \\[1em]
   a \sigma_6\sigma_9  , 
& b \sigma_6\sigma_9\sigma_{21} ,&  1
\end{pmatrix}
}
which can be close to Eq. (\ref{eq:MupFit}).
Let all singlet VEVs are real except $\sigma_9$ and $\sigma_{21}$ \cite{KimSeo11},
\dis{
\sigma_9=|\sigma_9| e^{i\theta  }\simeq|\sigma_9| e^{\frac{i\pi}{2} },~~\sigma_{21}=|\sigma_{21}|e^{i\phi  }.
}
The phase of $e^{i(\theta+\phi)}$ is fitted to the phase of
\dis{
 -  1.15\times 10^{-3} +(2.88\times 10^{-4} -0.534\times 10^{-4}t_5  ) e^{i\delta} \sim e^{i(\theta+\phi)}
}
In Table V, we list $\theta+\phi$ for a few $t_5$.  For $\delta_{\rm CKM}=\frac{\pi}{2}$ and $t_5\simeq 5.5$, we obtain $\phi\simeq -\theta$. Irrespective of the value of $\phi$, the CP phase in the Jarlskog determinent, $\delta_{\rm CKM}$, is the phase in $M_{u(31)}^w$ with the KS parametrization given in Eq. (\ref{eq:KSform}).
  %%%%%%%%%%%%%%%%%
\begin{table}[h!]
\begin{center}
\begin{tabular}{@{}lccc@{}} \toprule
 $t_5$ & $\frac{(-0.288+0.0534 t_5)}{1.15}$ &$\theta+\phi$ \\[0.1em] \colrule
 $0$  & $-0.250$  &    $-14.04^{\rm o}, -0.244\pi$ \\
  $5.5$   &$0.0054$  &    $0.286^{\rm o},\sim 0$ \\
 $10$   &$0.214$  &   $12.0^{\rm o},0.209\pi$ \\[0.2em]
  \botrule
\end{tabular} \label{tab:ab7} 
\end{center}
\caption{Phases of $M_{u(32)}^w$ for $\delta=\frac{\pi}{2}$. }
\end{table}

%%%%%%%%%%%%%%%%
%%%%%%%%%%%%%%%%
\subsection{CP phases in the quark and lepton sectors}\label{subsec:CPphase}
 
 As done before, let us diagonalize the symmetric fermion masses first. In the flipped SU(5) model, therefore, we diagonalize down-type quark masses and neutrino masses. Then, we consider up-type quarks and charged leptons. Then, the (3,1) elements of the mass matrices are the key.
For the third family members from $U$, masses of $t$ quark and $\tau$ lepton arise from
\dis{
 \tenb_{-1}(U)\leftrightarrow~ &\five_{+3}(U)\quad\qquad\qquad \times \five_{-2}(T_6),~ \left(t~ {\rm quark:} \sum_{{\rm Sect}~T_i}=6,\sum Q_{\rm anom}=+14 \right)\\[0.5em]
&\five_{+3}(U)\leftrightarrow\one_{-5}(U)~\times \fiveb_{+2}(T_6), \left(\tau~ {\rm lepton:} ~\sum_{{\rm Sect}~T_i}=6,\sum Q_{\rm anom}=-4 \right)\label{eq:ql31}
}

The phenomenologically determined leptonic mass element $M_{e(31)}^w$ can be obtained from Eq. (\ref{eq:Wminus}) by changing the quark parameters $\theta_i,\delta,\Delta, m_u, m_c, m_t$ to leptonic  parameters of Eq. (\ref{eq:PMNSpara}): $\Theta_i,\delta_L,\Delta_L, m_e, m_\mu, m_\tau$. Choose the leptonic $V$ matrix elements such that
\dis{
\frac{S_4}{C_4}=\frac{S_1}{C_1}, \frac{S_5}{C_5}=\frac{m_\mu}{m_\tau}\frac{S_2}{C_2}, S_6=0, \Delta_L=\delta_L.
}
Then, $M_{e(31)}^w/m_\tau\simeq -\sin\Theta_1\sin\Theta_2 e^{i\delta_L} $ where $\delta_L$ is the PMNS phase.

In our model, Table IV,  there are three $e^c$ fields in the leptonic case (instead of four $u^c$ fields in the quark case), and we can choose $S_{24}^A$ which  is the antisymmetric combination of $S_{24a}  $ and  $S_{24b}  $ in $T_4^0$. So, the leptonic mass matrix has four zero entries with the antisymmetric 1st row and antisymmetric 2nd column, 
\dis{
\begin{pmatrix} 
 0,& M_{e(12)}^w, &0\\
  M_{e(21)}^w, &0,&  M_{e(23)}^w \\
M_{e(31)}^w, &0,&  M_{e(33)}^w \\
 \end{pmatrix}
}
For the lepton phase, we need  $M_{e(31)}^w$ whose phase is $\delta_L$,\footnote{We have already presented $M_{u(31)}^w$ with phase $\delta_{\rm CKM}$.   We have this definite statement because we used the KS parametrization, Eqs. (\ref{eq:KSform}) and (\ref{eq:PMNSpara}),  of mixing angles \cite{KimSeo11}. } 
\dis{
M_{e(31)}^w: C_1( \five_U)S_{24}^S(\one_{T_4^0})H_d(\fiveb_{T_6}), \left( ~\sum_{{\rm Sect}~T_i}=10,\sum Q_{\rm anom}=-4 \right) 
}
where $S_{24}^S$ is the symmetric combination of $S_{24a}  $ and  $S_{24b}  $.  So, the product of FN singlets must satisfy 
\dis{
\sum_{{\rm Sect}~T_i}=2~{\rm mod.~}12,~\sum Q_{\rm anom}=+4 .
}
It is satisfied by $\sigma_5\sigma_6\sigma_9$ and $\sigma_2\sigma_{15}\sigma_{21}$.\footnote{In  $M_{u(31)}^w$, we did not include $\sigma_{10}$ in addition to  $\sigma_9$ for simplicity. Namely, it is equivalent to assuming $\langle \sigma_{10}\rangle=0$  or $ \sigma_{10}=| \sigma_{10}|e^{i\theta}$. In  $M_{e(31)}^w$ also, we consider only $\sigma_9$ for simplicity.} In our vacuum, we choose  $|\sigma_{21}|$ somewhat smaller than $|\sigma_{9}|$ but large enough to achieve a successful doublet-triplet splitting, viz. Eq. (\ref{eq:Svevs1}). Therefore, the phase $\delta_L$ of  $M_{e(31)}^w$ is mostly given by the phase of $\sigma_9$ as in  $M_{u(31)}^w$. In the leptonic case, this $\delta_L$ is    $\delta_{\rm PMNS}$.  But, for this interpretation to work, $\sigma_9$ should not appear in the neutrino mass matrix such that $U_{\rm PMNS}=U^\dagger_\nu U_e$ contains $\sigma_9$ only in $U_e$. The neutrino mass matrix is of the form $\five_{+3}M_\nu^w \five_{+3}$ which can arise from the following couplings,
 \dis{
\frac{1}{M^9} \, C_{3}[\five_{+3}(T_4^0)] &C_{3}[\five_{+3}(T_4^0)] H_u[\five_{-2}(T_6)]H_u[\five_{-2}(T_6)]
 C_{11}[\tenb_{-1}(T_3)]C_{11}[\tenb_{-1}(T_3)]\\
  &\cdot \sigma_3  [\one(T_4^0)] \sigma_3[\one(T_4^0)]\sigma_5[\one(T_6)] \sigma_5[\one(T_6)] \sigma_{21}[\one(T_1^0)]\sigma_{21}[\one(T_1^0)],\\
\frac{1}{M^7} \, C_{3}[\five_{+3}(T_4^0)] &C_{1}[\five_{+3}(U)] H_u[\five_{-2}(T_6)]H_u[\five_{-2}(T_6)]
 C_{11}[\tenb_{-1}(T_3)]C_{11}[\tenb_{-1}(T_3)]\\
   &\cdot \sigma_5[\one(T_6)] \sigma_6[\one(T_6)] \sigma_{21}[\one(T_1^0)]\sigma_{21}[\one(T_1^0)],\\
\frac{1}{M^8} \, C_{1}[\five_{+3}(U)] &C_{1}[\five_{+3}(U)] H_u[\five_{-2}(T_6)]H_u[\five_{-2}(T_6)]
 C_{11}[\tenb_{-1}(T_3)]C_{11}[\tenb_{-1}(T_3)]\\
 &\cdot \sigma_2 [\one(T_4^0)] \sigma_5[\one(T_6)] \sigma_6[\one(T_6)] \sigma_{21}[\one(T_1^0)]\sigma_{21}[\one(T_1^0)]\\
  \label{eq:nuMass}
 }
 where $\sigma_9$ does not appear. So, the phase in $\sigma_9$ is the PMNS phase. The generic magnitudes of masses from the above couplings are $(v_{\rm ew}^2/M)(V/M)^{8,6,7}$ where $V$ and $M$ are some scales around/above the GUT scale, and we can obtain reasonable strength for neutrino masses.
  
Equation (\ref{eq:ql31}) shows that the L-handed up-type quarks, appearing in $\tenb_{-1}$,  use charge lowering operators to couple to $W^-_\mu$ and  the L-handed charged leptons, appearing in $\five_{+3}$,  use charge raising  operators to couple to $W^+_\mu$. So,  we must consider the same charge charged-gauge boson $W_\mu$ to compare the signs of $\delta_{\rm CKM}$ and  $\delta_{\rm PMNS}$. Also, we must specify the signs of the effective Yukawa couplings in  $M_{u(31)}^w$ and  $M_{e(31)}^w$ dictated by string compactification. At this stage, we allow any sign for  $M_{u(31)}^w$ and  $M_{e(31)}^w$ since we considered only the selection rules.   If the signs of  $M_{u(31)}^w$ and  $M_{e(31)}^w$ are the same (opposite), then we conclude that  $\delta_{\rm CKM}$ and  $\delta_{\rm PMNS}$ have the opposite (same)  signs.\footnote{In the GG model \cite{GG74}, the  symmetric quark mass matrices are for neutrinos and up-type quarks.  The asymmetric quark mass matrices are for the down-type quarks and charged leptons via the same coupling $\tenb_0\five_0 H_{0\,d}$, and if we had tried the strategy we chose here  then we would have obtained the same sign for $\delta_{\rm CKM}$ and  $\delta_{\rm PMNS}$ irrespective of  the signs of  $M_{u(31)}^w$ and  $M_{e(31)}^w$. But this idea is not workable in the GG model because we lack an adjoint representation for breaking SU(5) down to the SM.} The case of opposite signs is consistent with the currently favored phases of  $\delta_{\rm CKM}$ \cite{KimSeo11}  and  $\delta_{\rm PMNS}$ \cite{T2K17}.

In the PS  type standard model SU(4)$\times$ SU(2)$_L\times$ SU(2)$_R$, we would have fermion matter spectra,  containing quark and lepton doublets,
\dis{
(\four, \two,\one)_{L}\oplus (\four ,\one, \two)_{R}+ \cdots
}
  Suppose that the Yukawa coupling $(\four, \two,\one)_{L} \times  (\four^*, \one,\two)_L\times  (\one, \two,\two)_h$ via Higgs $(\one, \two,\two)_h$ is present from the orbifold compactification. Then, the Yukawa coupling arises from the L-handed Higgs field doublets $\epsilon^{ij}(\one, \two,  ({ij}))_h=(\one, \two,(12))_h-(\one, \two,(21))_h$ where the R-hand index $(12)$ gives the Higgs doublet coupling to quark doublets and the R-hand index $(21)$ gives the Higgs doublet coupling to lepton  doublets. We use the same charge W, \ie $W^+_\mu$, for coupling to down-type quarks and charged leptons. So, the relative signs of  $M_{d(31)}^w$ and  $M_{e(31)}^w$ are  opposite if the product with FN singlet contributions give the same sign. If we use the mass matrices of  $M_{d(31)}^w$ and  $M_{e(31)}^w$ for asymmetric mass matrices as in the GG model, then   $\delta_{\rm CKM}$ and  $\delta_{\rm PMNS}$ have the opposite  signs. But, here one needs an example for breaking SO(10) down to  SU(4)$\times$ SU(2)$_L\times$ SU(2)$_R$, where the rank is not reduced, from the spectra of orbifold compactification. One may use the bulk fields for an adjoint representation as pointed out for $\Z_{6-II}$ in Ref. \cite{Raby05} and for $\Z_2\times \Z_2$ in Ref. \cite{Faraggi07} where the $N=2$ gauge multiplet  in an effective 5-dimensional SUSY model allows an adjoint representation of spin-0 fields.

%%%%%%%%%%%%%%%%
\section{Conclusion}\label{sec:Conclusion}

In this paper, we presented a theory toward understanding the quark and lepton mixing angles. Specifically, we presented a working example obtained from a string compactification \cite{Huh09} with $Q_{\rm anom}$ charge presented in \cite{KimKyaeNam17}. Explicit presentations were given for the CKM matrix.  The (3,3) element of quark mass matrix in the weak basis, is assumed to be close to the $t$-quark mass. Because there are only three L-handed quark doublets in the model, the up-type quark mass matrix is antisymmetric under the exchange of $a\leftrightarrow b$ among R-handed flavor indices  (or $u^c$ fields) obtained from $T_4^0$. This is because the multiplicity 2 for $\five_{-3}$ from $T_4^0$ is generic and there is no way to distinguish these two. The antisymmetric combination of $a$ and $b$ is named for the 1st family member  of $\five_{+3}$'s.  But, there are four  L-handed up-type quark doublets and the up-type quark families have a freedom to choose from these four. We used the freedom of choosing the unitary matrix for the R-handed quarks to fit to the data, and   showed that this model predicts reasonable mixing angles within experimental error bounds. Also, we studied the relation between   $\delta_{\rm CKM}$ and  $\delta_{\rm PMNS}$ by the phases of some SM singlet scalar fields, assuming that all Yukawa coupling constants from string compactification are real. For the proton decay problem, a $\Z_2$ matter parity cannot be introduced consistently with the solution of the doublet-triplet splitting problem by the GUT scale VEVs, $\langle \tenb_{-1}(T_3)\rangle $ and $\langle \ten_{+1}(T_9)\rangle $. But, we showed that the proton decay operator appears at a dimension 10 level, which can be made small enough while achieving the doublet-triplet splitting. It will be interesting if a kind of R parity is found within the scheme, which will be pursued in the future.

 %%%%%%%%%%%%%%%%%%%%%%%%%%%%%%%%%%%%%%%%%%%%%%%%%%%%%%%%%%%%%%%%%%%%
\acknowledgments{I have benefitted from discussions with S. Kim, B. Kyae, and S. Nam. This work is supported in part by the National Research Foundation (NRF) grant  NRF-2015R1D1A1A01058449  and by the IBS (IBS-R017-D1-2014-a00). }

  %%%%%%%%%%%%%%%%%%  

\end{document}